\documentclass[twocolumn,twocolappendix]{aastex631}  
\usepackage{amsmath}
\usepackage{natbib}
\usepackage{booktabs}
\usepackage{xcolor}
\usepackage{float}
\newcommand{\arcsecond}{^{\prime\prime}}

\usepackage{bm}
\usepackage{esvect}

\usepackage{hyperref}
\hypersetup{
    colorlinks=true, 
    linkcolor=blue, 
    citecolor=blue, 
    urlcolor=blue 
}
\begin{document} 
    \title{Ten Aligned Orbits: Planet Migration in the Era of JWST and Ariel}
    \shorttitle{Ten aligned planetary systems}
    
   \author{J. Zak}
   \affiliation{Astronomical Institute of the Czech Academy of Sciences, Fri\v{c}ova 298, 25165 Ond\v{r}ejov, Czech Republic}

    \author{H.\,M.\,J. Boffin}
   \affiliation{European Southern Observatory, Karl-Schwarzschild-str. 2, 85748 Garching, Germany}

   \author{A. Bocchieri}
   \affiliation{Dipartimento di Fisica, La Sapienza Università di Roma, Piazzale Aldo Moro 5, Roma, 00185, Italy}

      \author{E. Sedaghati}
   \affiliation{European Southern Observatory, Casilla 13, Vitacura, Santiago, Chile}

      \author{Z Balkoova}
   \affiliation{Astronomical Institute of the Czech Academy of Sciences, Fri\v{c}ova 298, 25165 Ond\v{r}ejov, Czech Republic}

      \author{P. Kabath}
   \affiliation{Astronomical Institute of the Czech Academy of Sciences, Fri\v{c}ova 298, 25165 Ond\v{r}ejov, Czech Republic}

   \correspondingauthor{Jiri Zak}
   \email{zak@asu.cas.cz}

   

\begin{abstract}
Understanding the diverse formation and migration pathways that shape exoplanetary systems requires characterizing both their atmospheric properties and their orbital dynamics. A key dynamical diagnostic is the projected spin-orbit angle—the alignment between the stellar spin and the planetary orbit—which provides crucial tests for theoretical models. This angle can be determined using the Rossiter-McLaughlin effect. Although measurements exist for over 200 planets, the overall distribution of these angles is not fully understood, motivating further observations across the full parameter space. We analyze archival HARPS and HARPS-N spectroscopic transit time series of nine gas giant exoplanets on short orbits and one brown dwarf. We derive their projected spin-orbit angle $\lambda$.  We find aligned projected orbits for all nine gas giants as well as the brown dwarf. Furthermore, we are able to derive the true spin-orbit angle for the brown dwarf EPIC 219388192b, $\psi = $25$^{+11}_{-14}$\,deg. These projected prograde orbits are consistent with quiet disc migration disfavoring violent events exciting the orbits in the history of these systems. Finally, we investigate the current overlap between spin-orbit angle measurements and atmospheric characterization targets. While we find no strong observational biases due to the spin-orbit angle, we note that the majority of planets with atmospheric data still lack spin-orbit measurements. This incompleteness of the dynamical information may limit the interpretation of upcoming atmospheric surveys.

\end{abstract}


\section{Spin-orbit angle of exoplanets} 
\label{sec:intro}

Current theories infer that hot and warm Jupiters can be formed via three possible scenarios: in-situ formation, disc migration, and disc-free migration \citep{daw18,iko25}. However, the occurrence rates of each of these planetary migration pathways are still unconstrained by observations. 
Currently, we can study the past evolution of giant planets using two complementary approaches. 

The first approach is to study their dynamical parameters \citep{bi13,clo24} such as the planetary masses, eccentricities, orbital periods and absolute inclinations\footnote{Relative inclinations are not physical parameters as they depend on the line of sight.} (spin-orbit angles).
The spin-orbit angle of an exoplanet, that is the angle between the stellar spin axis and the normal to the orbital plane, has emerged as a powerful probe of exoplanetary dynamics \citep{quel00,fab09,alb22}, although usually only its sky-projected value is measured using the Rossiter-McLaughlin (R-M) effect. Thanks to the availability of more precise high-resolution spectrographs, recent years have seen a surge in the numbers of their detections \citep[e.g.,][]{esp24,knud24,zak25}. Measurements of gas giants have revealed diverse configurations ranging from perfectly aligned and prograde orbits to polar and retrograde configurations. Even compact systems with multiple transiting planets have shown a large diversity in orbital architectures with systems having planets on almost perpendicular orbits \citep{bo21,yu25}, two highly misaligned planets around a star in a triple system \citep{hjo21}, or even with a planet in perpendicular orbit around a binary brown dwarf \citep{bay25}.

Constraining evolutionary pathways only from the dynamical parameters can sometimes be degenerate as distinct evolutionary histories can produce the same orbital configuration. Furthermore, the tidal forces remove dynamical information with time as they tend to circularize and realign orbits \citep{spal22}. This is especially true for massive planets on short orbits around cool stars that display a high degree of alignment \citep{rus24}. Hence, planets that are not significantly affected by the tidal forces (e.g., warm Jupiters) represent more amenable targets to probe the evolutionary pathways. Currently, there are over 250 measurements of the projected spin-orbit angle, but clear identification and description of trends in the exoplanetary population is still limited, highlighting the need for additional measurements \citep{knud24}.

The second approach to investigate the history of planetary systems is to study the composition of their atmospheres. The elemental abundance ratios such as C/O, N/O and others can constrain evolutionary pathways \citep{ob11, tur21}. More specifically, the planets undergoing quiet disc migration are predicted to accrete material (e.g., oxygen-rich planetesimals) changing their composition with respect to the planets moving inward with disc-free migration after the disc dispersal. However, even this approach is not unambiguous as disequilibrium chemistry processes such as photochemistry and tidal heating can change the chemical composition of the atmosphere, partially erasing the record of its past migration \citep{for20,tsai23}. 

Combining these two approaches on a population level allows identifying possible trends and biases. This was already recognized in the ongoing BOWIE-ALIGN survey targeting four aligned and four misaligned planets to determine their composition with JWST \citep{kirk24}. In the future, the \textit{Ariel} mission \citep{tin18}  will provide atmospheric measurements for about a thousand exoplanets. As one of the main science questions\footnote{\url{https://sci.esa.int/documents/34022/36216/Ariel_Definition_Study_Report_2020.pdf}} is to investigate ``How do planets and planetary systems form and evolve?", the \textit{Ariel} consortium has recently established the stellar obliquity sub-working group to obtain spin-orbit angle measurements for all its potential targets. Having the spin-orbit angle provides highly complementary information for determining the plausible evolutionary scenarios.

In this work, we provide measurements of the projected spin-orbit angle of nine gas giants and one brown dwarf. Furthermore, we assess the available spin-orbit angle measurements of planets targeted by HST and/or JWST and provide an outlook for the \textit{Ariel} mission.

\section{Data sets and their analyses} \label{sec:data}
In this study, we have mostly focused on Jupiters on short orbits having archival spectroscopic data.
Our datasets come from two sources: three datasets (WASP-35, WASP-162 and Qatar-7) come from the HARPS-N instrument \citep{cos12} mounted at the 3.58-m TNG telescope at La Palma, Canary Islands. The rest of the targets come from the HARPS instrument \citep{ma03} mounted at the ESO 3.6-m telescope at La Silla, Chile. The HARPS data were obtained from the ESO science archive\footnote{\url{http://archive.eso.org}} (program IDs 0102.C-0618(A), 089.C-0151(B), 094.C-0090(A), 095.C-0105(A) and 105.20B1.003), while the HARPS-N data (program IDs CAT18A\_138, CAT18B\_125) were obtained from the TNG archive\footnote{\url{http://archives.ia2.inaf.it/tng/}}. Table \ref{obs_logs} shows the properties of the data sets that were used. The downloaded data are fully reduced products as processed with the HARPS Data Reduction Software (DRS version 3.8). Each spectrum is provided as a merged 1D spectrum resampled onto a 0.001 nm uniform wavelength grid. The wavelength coverage of the spectra spans from 380 to 690 nm, with a resolving power of R\,$\approx$\,115\,000, corresponding to 2.7~km\,s$^{-1}$ per resolution element. The spectra are already corrected to the Solar system barycentric frame of reference. We obtained the radial velocity measurements acquired from the HARPS and HARPS-N DRS pipelines together with their uncertainties. The radial velocity data for the WASP-35 and Qatar-7 systems obtained with HARPS-N were showing unknown systematics that hindered further analysis. Hence, we have decided to obtain the radial velocities through the Yabi\footnote{\url{https://github.com/muccg/yabi/}}  workflow interface \citep{hunter12}, which is maintained by the Italian centre for Astronomical Archive (IA2)\footnote{\url{https://www.ia2.inaf.it/}}. As both stars are F-type stars we have used dedicated F-type star line mask to retrieve the radial velocities in the same format as the rest of the HARPS-N pipeline.  We further summarize the properties of the studied systems in Table \ref{tab:targets1}.

\begin{deluxetable*}{lcccccc}
\tablecaption{Observing logs for all ten systems. The number in parenthesis represents the number of frames taken in-transit.\label{obs_logs}}
\tablehead{
\colhead{Target} & \colhead{Night} & \colhead{No.} & \colhead{Exp.} & \colhead{Airmass} & \colhead{Median} & \colhead{Instrument} \\
\colhead{} & \colhead{Obs} & \colhead{frames} & \colhead{Time (s)} & \colhead{range} & \colhead{S/N$^\mathrm{a}$} & \colhead{} 
}
\startdata
WASP-35 & 2017-11-18/19 & 39 (18) & 600 & 3.31-1.22-1.48 & 6-34 & HARPS-N \\
WASP-44 & 2018-11-26/27 & 17 (8) & 900 & 1.05-2.14 & 7-18 & HARPS \\ 
WASP-45 A & 2018-11-03/04 & 20 (7) & 900 & 1.01-1.80 & 26-13 & HARPS \\
WASP-54 A & 2012-05-09/10 & 38 (27) & 600 & 1.68-1.14-1.69 & 26-43 & HARPS \\
WASP-91 & 2018-11-01/02 & 31 (13) & 900 & 1.33-1.32-1.67 & 21-28 & HARPS \\
WASP-99 & 2014-10-18/19 & 22 (18) & 900 & 1.11-1.07-1.59 & 21-94 & HARPS \\
WASP-129 A & 2015-04-23/24 & 23 (10) & 900 & 1.15-1.03-1.39 & 20-29 & HARPS \\
WASP-162 & 2018-04-02/03 & 36 (24) & 600 & 2.40-1.45-2.50 & 10-17 & HARPS-N \\
Qatar-7 & 2018-11-10/11 & 21 (13) & 900 & 1.04-1.01-1.58 & 9-13 & HARPS-N \\
EPIC 219388192 & 2021-08-04/05 & 24 (10) & 900-1100 & 1.15-1.02-2.34 & 7-15 & HARPS \\
\enddata \tablenotetext{a}{S/N in the extracted 1-D spectrum, per spectral resolution element in the order at 550 nm.}
\end{deluxetable*}

\setlength{\tabcolsep}{6pt} 

\begin{splitdeluxetable*}
{lcccccBlccccc}

\tablewidth{0pt} 
\tablecaption{Properties of the targets (star and companion).\label{tab:targets1}}
\tablehead{
\colhead{Parameters} & \colhead{WASP-35} & \colhead{WASP-44} & \colhead{WASP-45 A} & 
\colhead{WASP-54 A} & \colhead{WASP-91} &\colhead{Parameters}&\colhead{WASP-99} & 
\colhead{WASP-129 A} & \colhead{WASP-162} & 
\colhead{Qatar-7} & 
\colhead{EPIC 219388192} 
}
\startdata
Star& & && & & Star \\
V mag & 10.95 & 12.9 & 12.0 & 11.3 & 12.0 &V mag & 9.5 & 12.3 $\pm$ 0.6 & 12.2 & 13.03 & 12.54 \\
Sp. Type & F9 & G8 & K2 & F9 & K3 &Sp. Type & F8 & G1 & K0 & F4 & G1 \\
M$_{\mathrm{s}}$ (M$_\odot$) & 1.10 $\pm$ 0.08 & 0.95 $\pm$ 0.08 & 0.95 $\pm$ 0.10 & 1.15 $\pm$ 0.09 & 0.84 $\pm$ 0.07 &M$_{\mathrm{s}}$ (M$_\odot$) & 1.48 $\pm$ 0.10 & 1.00 $\pm$ 0.03 & 0.95 $\pm$ 0.04 & 1.409 $\pm$ 0.026 & 1.101 $^{+0.039}_{-0.045}$ \\
R$_{\mathrm{s}}$ (R$_\odot$) & 1.09 $\pm$ 0.14 & 0.90 $\pm$ 0.22 & 1.00 $\pm$ 0.25 & 1.40 $\pm$ 0.19 & 0.86 $\pm$ 0.03 &R$_{\mathrm{s}}$ (R$_\odot$) & 1.76 $\pm$ 0.20 & 0.90 $\pm$ 0.02 & 1.11 $\pm$ 0.05 & 1.564 $\pm$ 0.021 & 1.029 $\pm$ 0.016 \\
T$_{\mathrm{eff}}$ (K) & 6050 $\pm$ 100 & 5400 $\pm$ 150 & 5100 $\pm$ 200 & 6100 $\pm$ 100 & 4920 $\pm$ 80 &T$_{\mathrm{eff}}$ (K) & 6150 $\pm$ 100 & 5900 $\pm$ 100 & 5300 $\pm$ 100 & 6311 $\pm$ 50 & 5755 $\pm$ 49 \\
$v\,\mathrm{sin}i_*$ (km/s) & 2.4 $\pm$ 0.6 & 3.2 $\pm$ 0.9 & 2.3 $\pm$ 0.7 & 4.0 $\pm$ 0.8 & 2.4 $\pm$ 0.4 & $v\,\mathrm{sin}i_*$ (km/s) & 6.8 $\pm$ 0.5 & 2.7 $\pm$ 0.6 & 1.0 $\pm$ 0.8 & 13.8 $\pm$ 0.5 & 5.6 $\pm$ 0.5  \\
\hline
Companion& && & & & Companion \\
M$_{\mathrm{p}}$ (M$_{\mathrm Jup}$) & 0.72 $\pm$ 0.06 & 0.889 $\pm$ 0.062 & 1.007 $\pm$ 0.053 & 0.626 $\pm$ 0.023 & 1.34 $\pm$ 0.08 &M$_{\mathrm{p}}$ (M$_{\mathrm Jup}$) & 2.78 $\pm$ 0.13 & 1.0 $\pm$ 0.1 & 5.2 $\pm$ 0.2 & 1.88 $\pm$ 0.25 &39.21$^{+0.91}_{-1.1}$ \\
R$_{\mathrm{p}}$ (R$_{\mathrm Jup}$) & 1.32 $\pm$ 0.03 & 1.14 $\pm$ 0.11 & 1.16$^{+0.28}_{-0.14}$ & 1.65$^{+0.09}_{-0.18}$ & 1.03 $\pm$ 0.04 &R$_{\mathrm{p}}$ (R$_{\mathrm Jup}$) & 1.10 $\pm$ 0.08 & 0.93 $\pm$ 0.03 & 1.00 $\pm$ 0.05 & 1.70 $\pm$ 0.03 &0.941 $\pm$ 0.19 \\
Period (d) & 3.161575(2) & 2.4238039(87) & 3.1260876(35) & 3.693649(13) & 2.798581(3) &Period (d) & 5.75251(4) & 5.748145(4) & 9.62468(1) & 2.032046(97) &5.292600(18) \\
$\mathrm{T_0}$ - 2450000 (d) & 5531.47907(15) & 5434.37600(40)  & 5441.26925(58) & 5522.04373(79) & 6297.7190(2)  &$\mathrm{T_0}$ - 2450000 (d) &  6224.9824(14) & 7027.4373(2) & 7701.3816(6) & 8043.32075(16) & 7341.037107(90)\\
a (au) & 0.04317(28) & 0.03473(41) & 0.04054(90) & 0.0497(5) & 0.037 $\pm$ 0.001 &a (au) & 0.0717 $\pm$ 0.0016 & 0.0628 $\pm$ 0.0007 & 0.0871 $\pm$ 0.0013 & 0.0352 $\pm$ 0.0002 & 0.06206$^{(+71)}_{(-84)}$  \\
e & 0 & 0 & 0 & 0 & 0 &e & 0 & 0 & 0.434 $\pm$ 0.005 & 0 & 0.1891 $\pm$ 0.002 \\
i (deg) & 87.96$^{+0.31}_{-0.25}$ & 86.02$^{+1.11}_{-0.86}$ & 84.47$^{+0.54}_{-0.79}$ & 84.8 $\pm$ 1.6 & 86.8 $\pm$ 0.4 &i (deg) & 88.8 $\pm$ 1.1 & 87.7 $\pm$ 0.2 & 89.3 $\pm$ 0.5 & 89.0 $\pm$ 1.0 & 88.53$^{+0.24}_{-0.27}$ \\
T$_{\mathrm{eq}}$ (K) & 1450 $\pm$ 20 & 1343 $\pm$ 64 & 1198 $\pm$ 69 & 1742 $\pm$ 69 & 1160 $\pm$ 30 &T$_{\mathrm{eq}}$ (K) & 1480 $\pm$ 40 & 1100 $\pm$ 25 & 910 $\pm$ 20 & 2053 $\pm$ 15 &1164 $\pm$ 40 \\
Discovery ref. & Enoch et al. (2011) & Anderson et al. (2012) & Anderson et al. (2012) & Faedi et al. (2013) & Anderson et al. (2017) &Discovery ref. & Hellier et al. (2014) & Maxted et al. (2016) & Hellier et al. (2019) & Alsubai et al. (2019) & Curtis et al. (2016) \\
Prior ref. & Bai et al. (2022) & Anderson et al. (2012) & Anderson et al. (2012) & Saha (2024) & Maciejewski (2022) &Prior ref. & Hellier et al. (2014) & Maxted et al. (2016) & Hellier et al. (2019) & Appendix A & Carmichael et al. (2019) \\
\enddata
\end{splitdeluxetable*}

\subsection{Rossiter-McLaughlin effect}

\begin{figure*}
\includegraphics[width=0.45\textwidth]{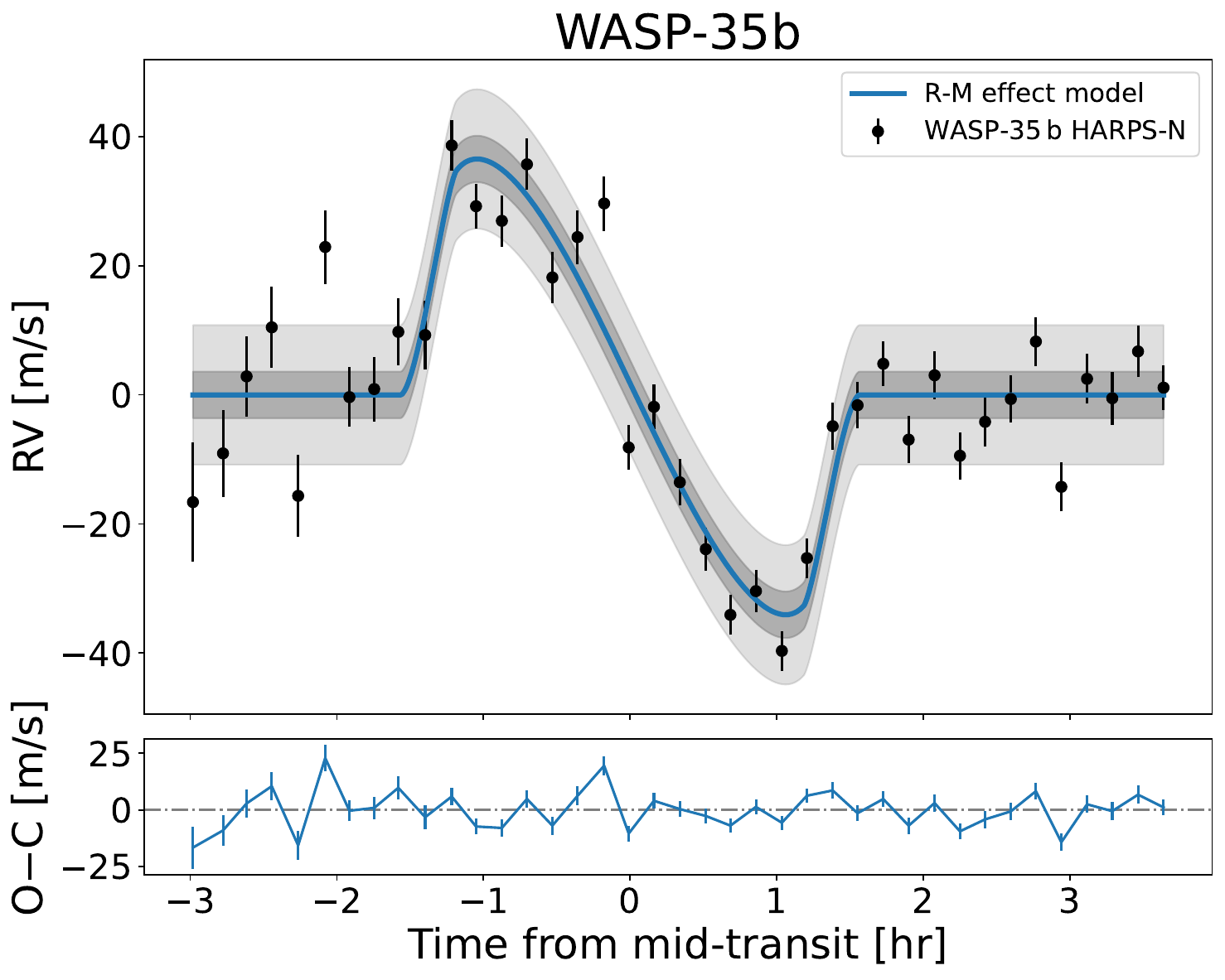}\includegraphics[width=0.45\textwidth]{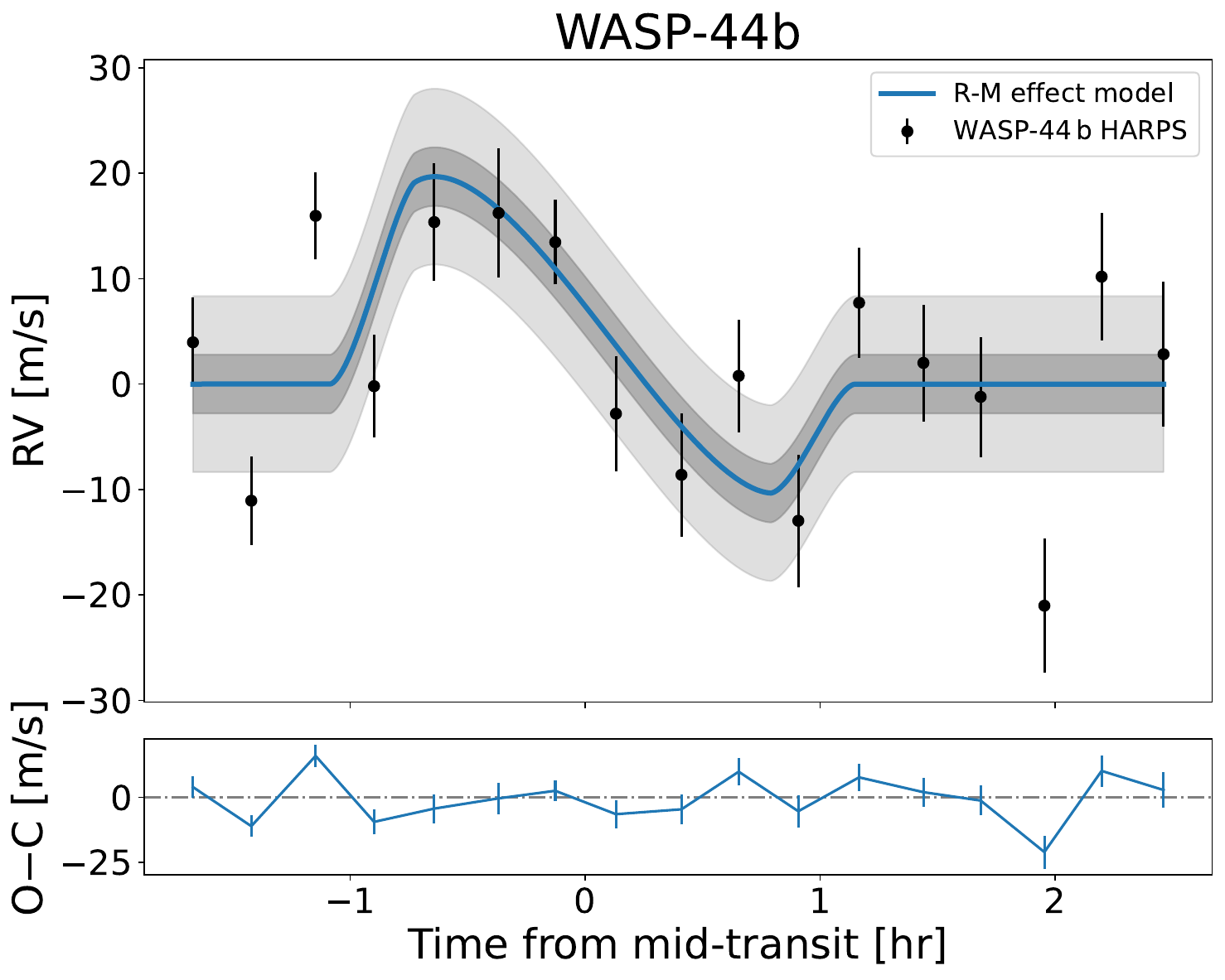}\\
\includegraphics[width=0.45\textwidth]{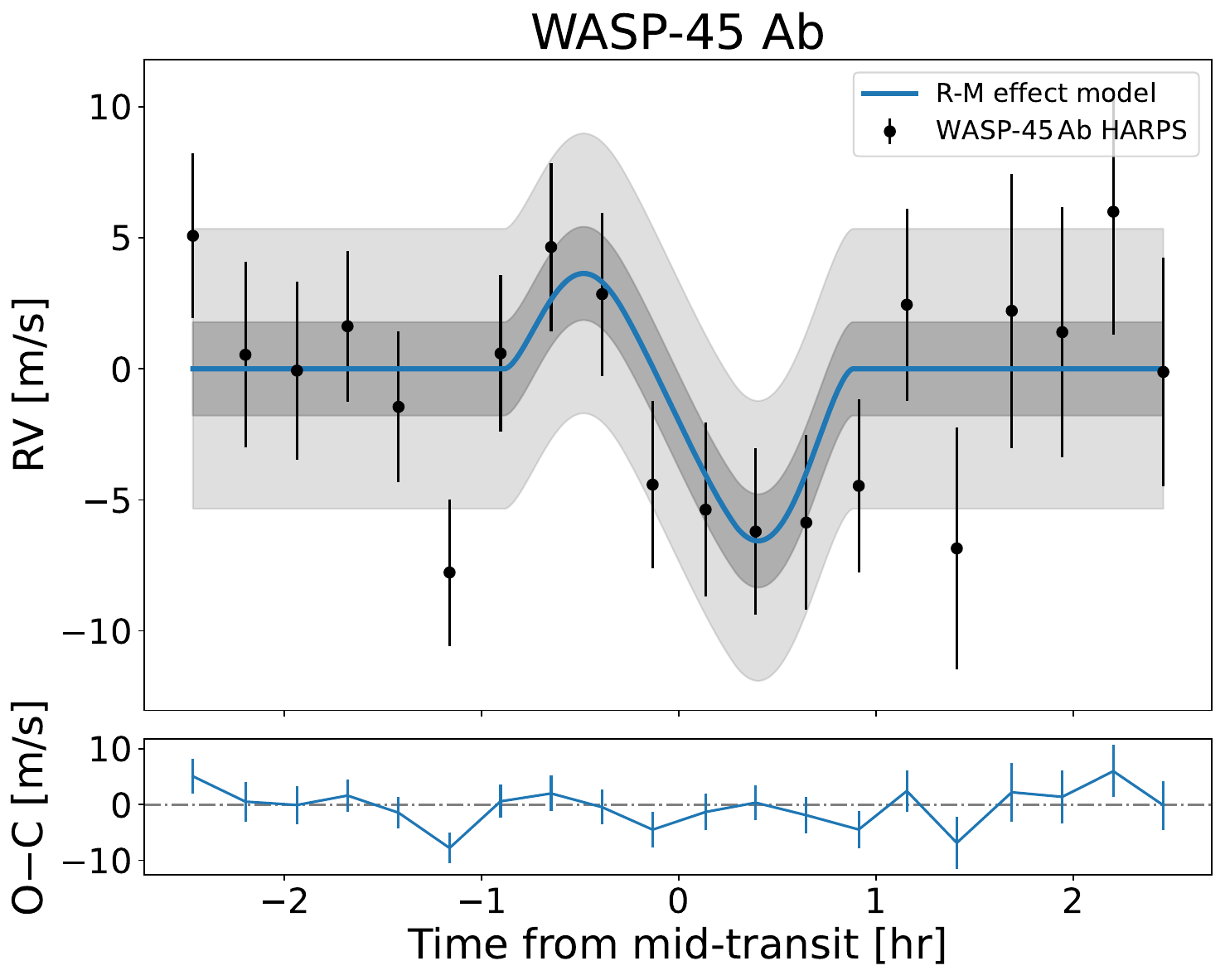}\includegraphics[width=0.45\textwidth]{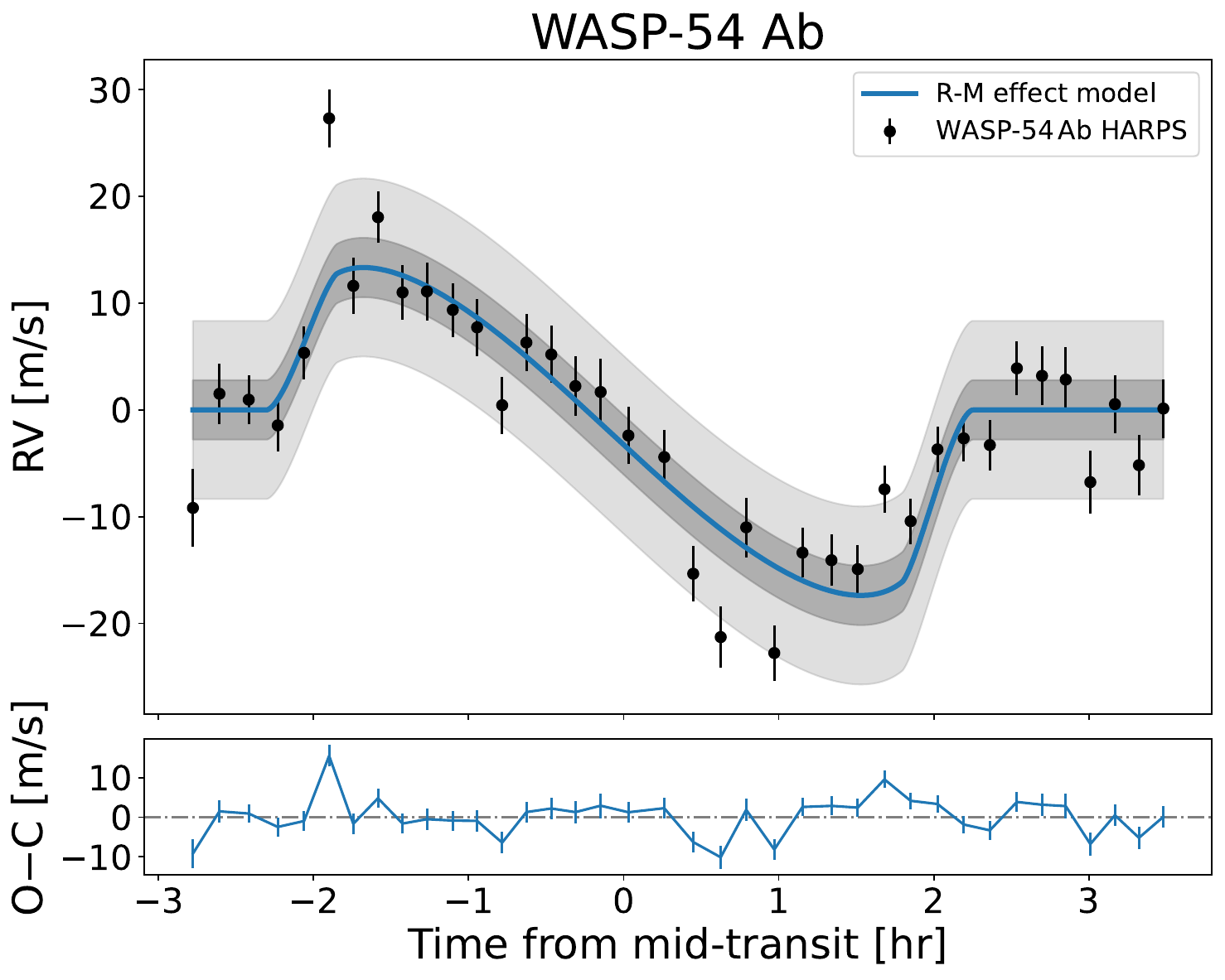}\\
\includegraphics[width=0.45\textwidth]{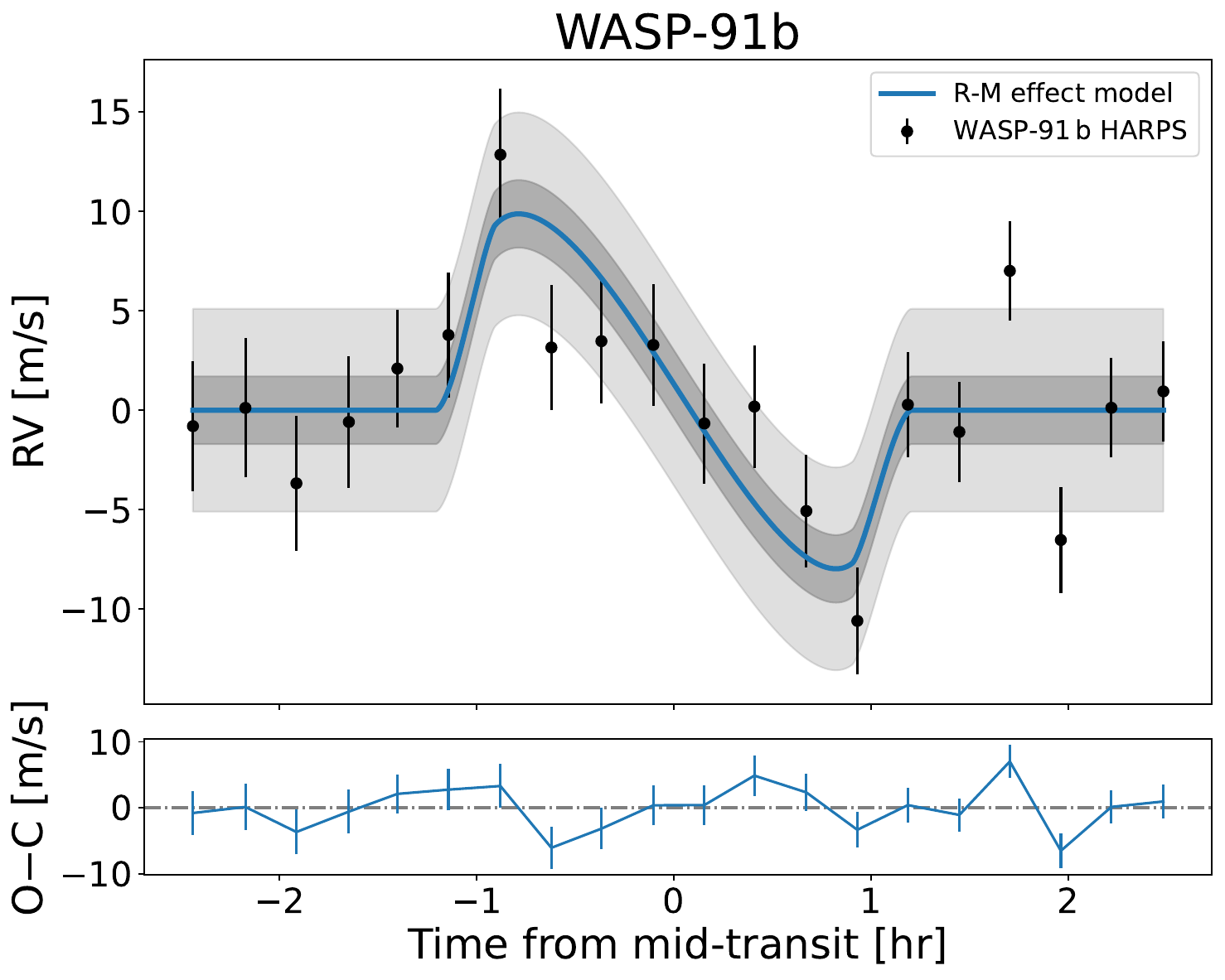}\includegraphics[width=0.45\textwidth]{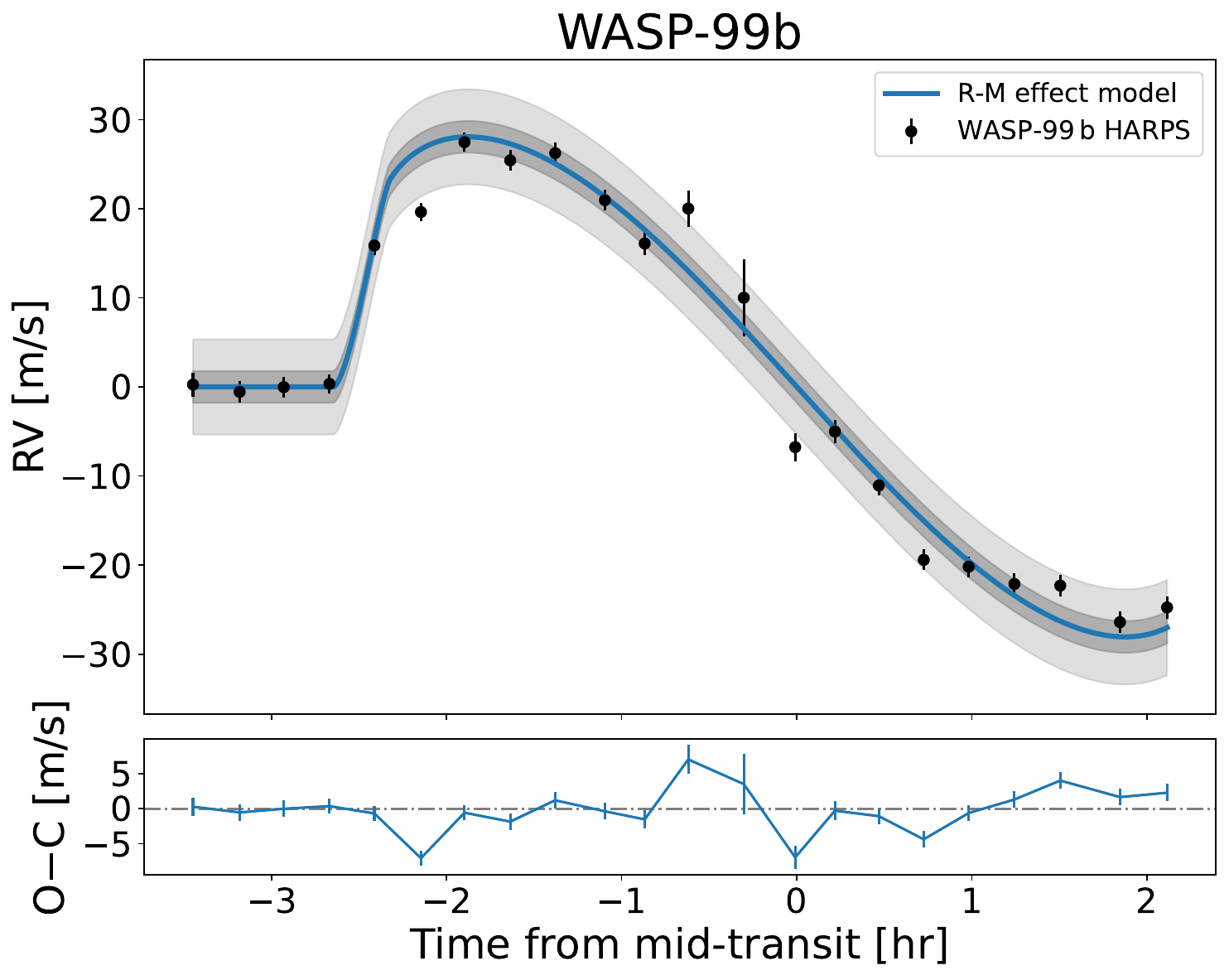}
\caption{The Rossiter-McLaughlin effect of WASP-35b, WASP-44b, WASP-45\,Ab, WASP-54\,Ab, WASP-91b, and WASP-99b. The observed data points (black) are shown with their error bars. The systemic and Keplerian orbit velocities were removed. The blue line shows the best fitting model to the data, together with 1-$\sigma$ (dark grey) and 3-$\sigma$ (light grey) confidence intervals.}
\label{f:35}
\end{figure*}

\begin{figure*}
\includegraphics[width=0.45\textwidth]{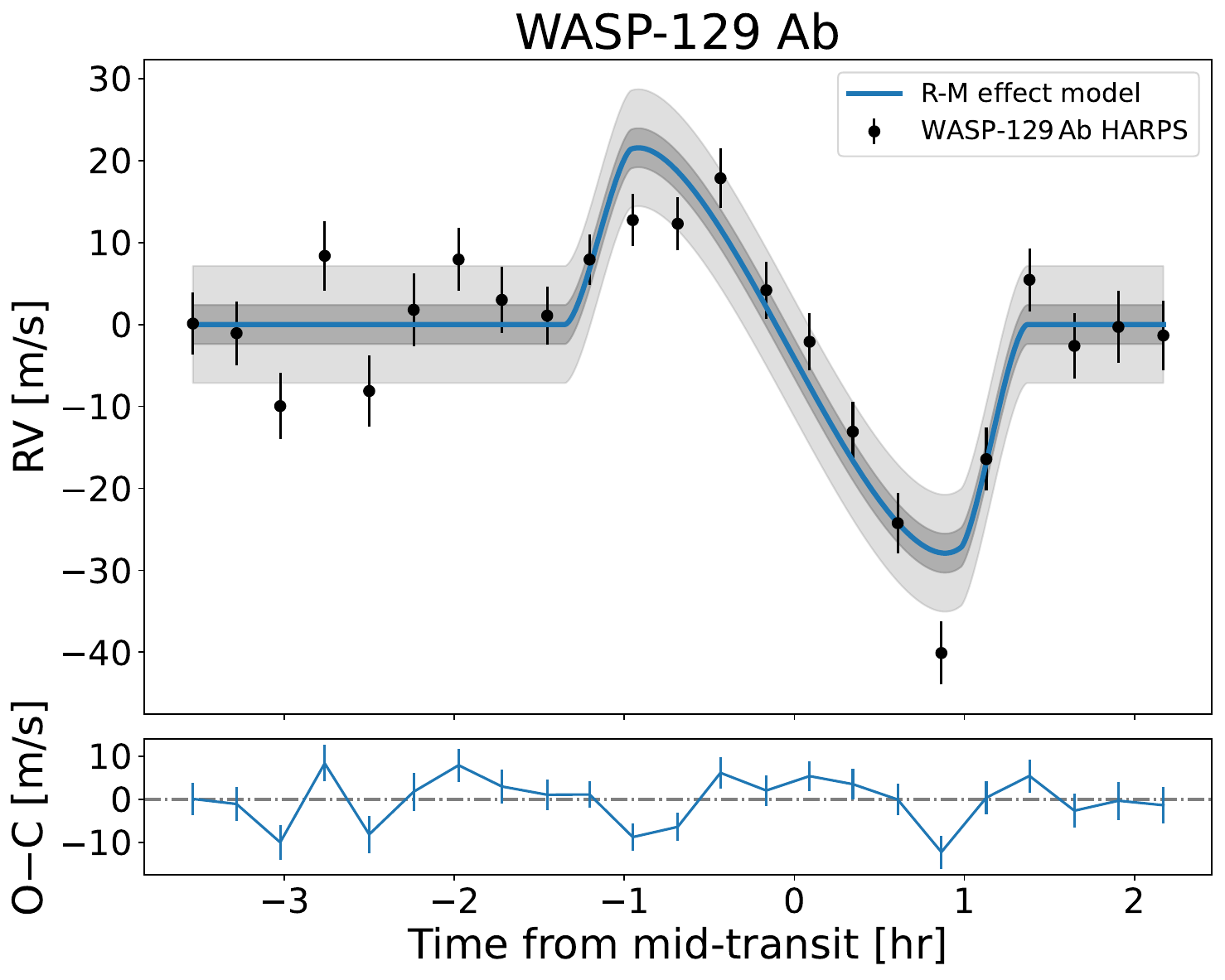}\includegraphics[width=0.45\textwidth]{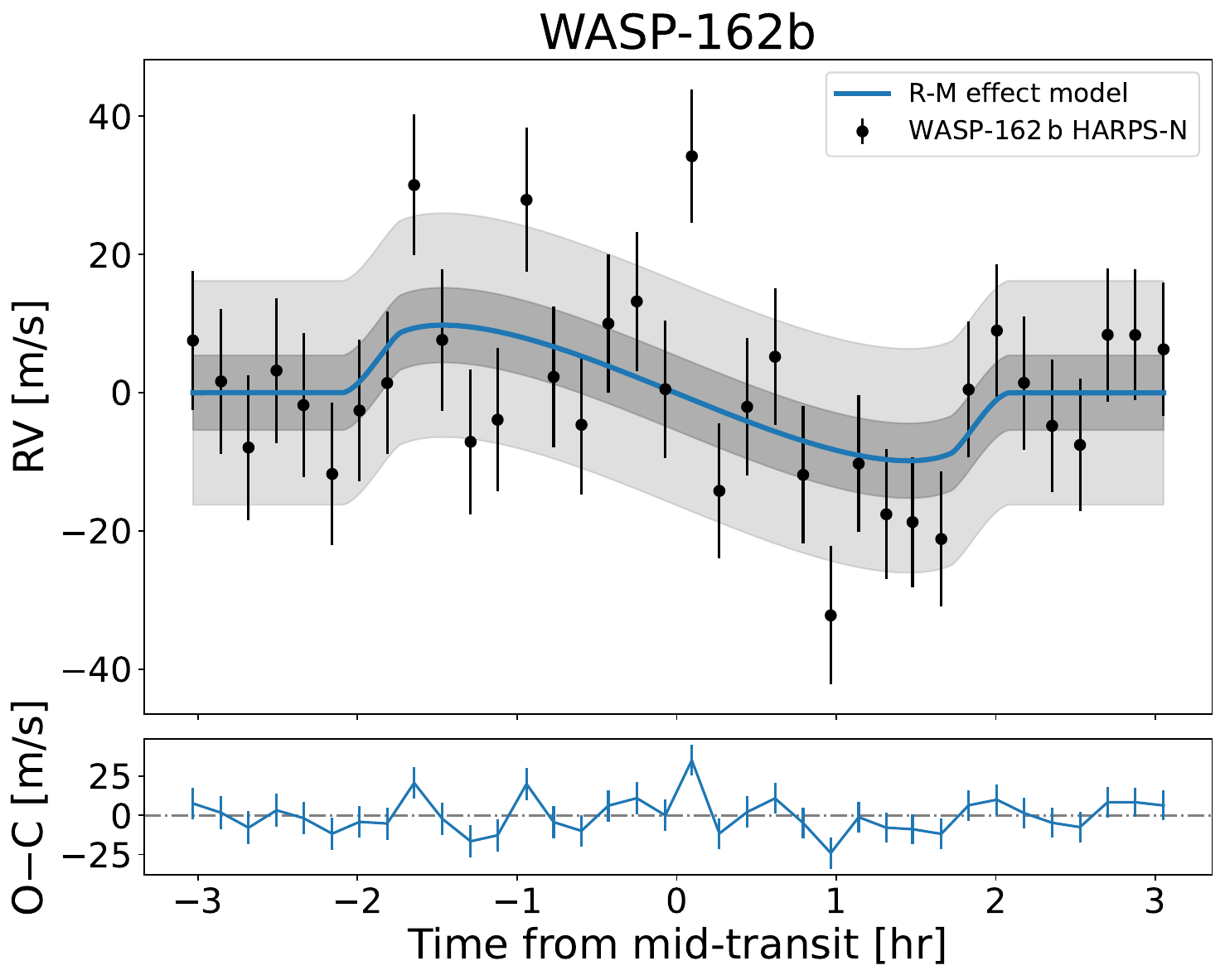}\\
\includegraphics[width=0.45\textwidth]{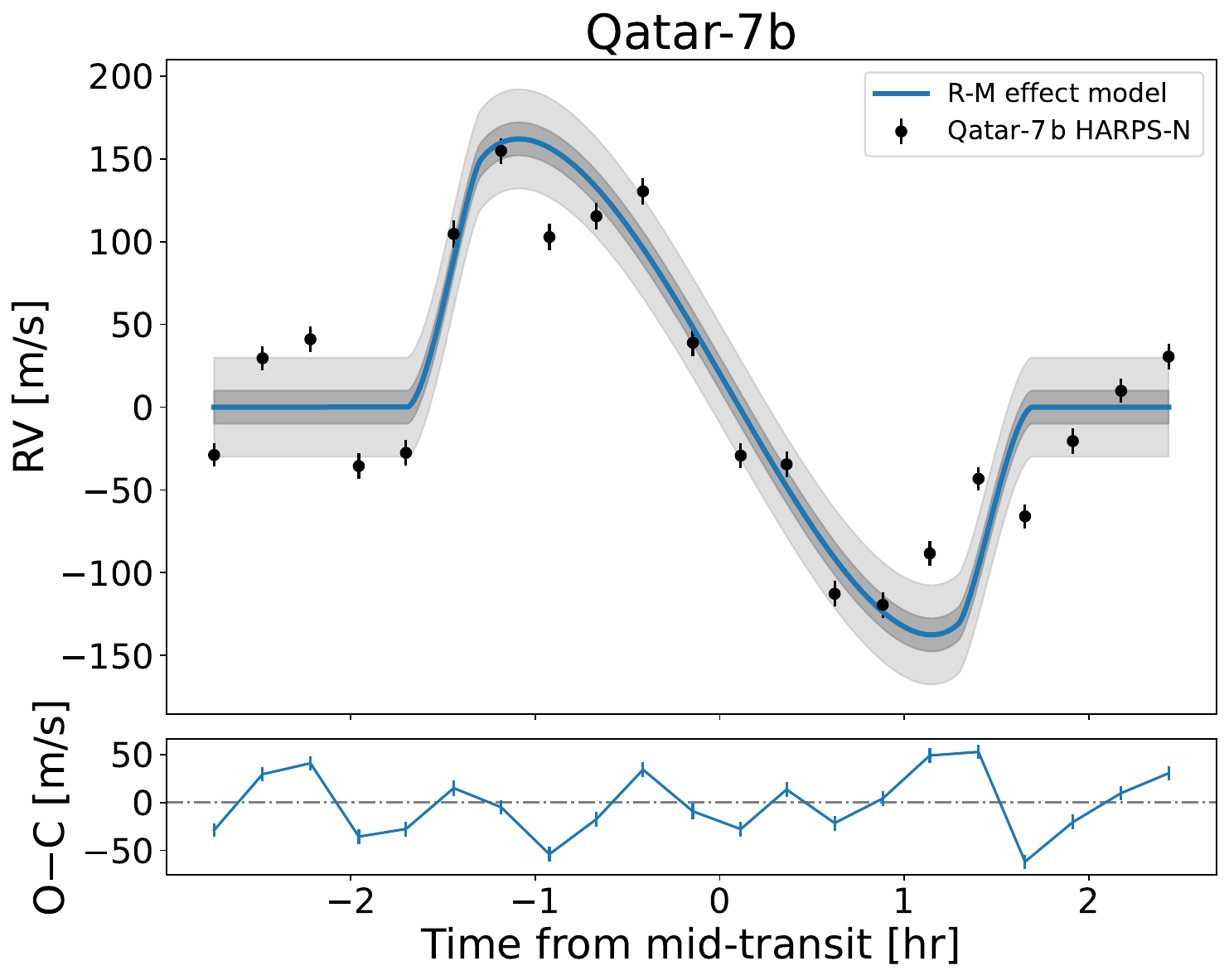}\includegraphics[width=0.45\textwidth]{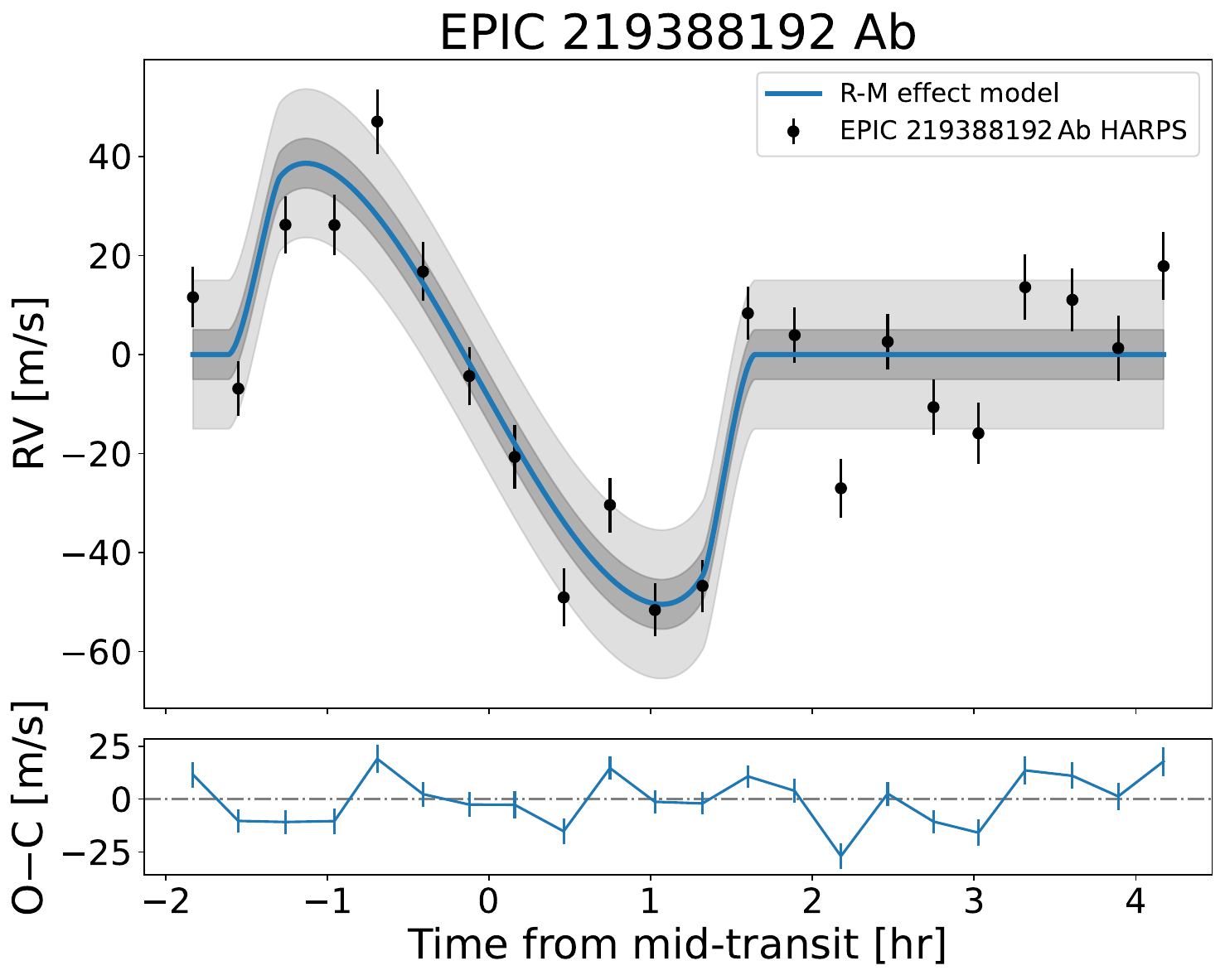}
\caption{Same as Fig.~\ref{f:35} for WASP-129\,Ab, WASP-162b, Qatar-7b, and EPIC 219388192b.}
\label{f:129}
\end{figure*}

The R-M effect causes a spectral line asymmetry, or, equivalently, an asymmetry in the cross-correlation function (CCF) that manifests as anomalous radial velocities during the transit of the exoplanet. To measure the projected spin-orbit angle of the system ($\lambda$), we followed the methodology of \citet{zak24a}: we fitted the RVs with a composite model, which includes a Keplerian orbital component as well as the R-M anomaly. This model is implemented in the \textsc{ARoMEpy}\footnote{\url{https://github.com/esedagha/ARoMEpy}} \citep{seda23} package, which utilizes the \textsc{Radvel} Python module \citep{ful18} for the formulation of the Keplerian orbit. \textsc{ARoMEpy} is a Python implementation of the R-M anomaly described in the \textsc{ARoME} code \citep{bou13}. We used the R-M effect function defined for RVs determined through the cross-correlation technique in our code. We set Gaussian priors for the RV semi-amplitude ($K$) and the systemic velocity ($\Gamma$). In the R-M effect model, we fixed the following parameters to values reported in the literature: the orbital period ($P$), the planet-to-star radius ratio ($R_{\rm{p}}/R_{\rm{s}}$), and the eccentricity ($e$). The parameter $\sigma$, which is the width of the CCF and represents the effects of the instrumental and turbulent broadening, was measured on the data and kept fixed. Furthermore, we used the ExoCTK\footnote{\url{https://github.com/ExoCTK/exoctk}} tool to compute the quadratic limb-darkening coefficients with ATLAS9 model atmospheres \citep{cas03} in the wavelength range of the HARPS and HARPS-N instruments (380-690\,nm). We set Gaussian priors using the literature value and uncertainty derived from transit modeling (Tab.~\ref{tab:targets1}) on the following parameters during the fitting procedure: the central transit time ($T_C$), the orbital inclination ($i$), and the scaled semi-major axis ($a/R_{\rm{s}}$). Uniform or Gaussian priors were set on the projected stellar rotational velocity ($\nu\,\sin i_*$). Uniform priors were set on the sky-projected angle between the stellar rotation axis and the normal of the orbital plane ($\lambda$). To obtain the best fitting values of the parameters, we employed three independent Markov chain Monte Carlo (MCMC) ensemble simulations using the \textsc{Infer}\footnote{\url{https://github.com/nealegibson/Infer}} implementation with the Affine-Invariant Ensemble Sampler. We initialized the MCMC at parameter values found by the Nelder-Mead method perturbed by a small value. We used 20 walkers each with 12\,500 steps, burning the first 2\,500. As a convergence check, we ensured that the Gelman-Rubin statistic \citep{gel92} is less than 1.001 for each parameter. Using this setup we obtain our results and present them in Sect.~\ref{sec:resrm} and in Figs.~\ref{f:35} to \ref{f:129}.

\section{Results} \label{sec:res}

\subsection{Projected stellar spin-orbit angle measurement} \label{sec:resrm}

We have measured the projected spin-orbit angle ($\lambda$) of ten targets for the first time: WASP-35b, WASP-44b, WASP-45 Ab, WASP-54 Ab, WASP-91b, WASP-99b, WASP-129 Ab, WASP-162 b, Qatar-7b and EPIC 219388192\,Ab. We provide their brief description below.

\textbf{WASP-35b} is an inflated hot Jupiter orbiting a F9V host star on a short 3.2-day orbit \citep{eno11}. 
In our analysis, we infer an aligned orbit of WASP-35b with $\lambda =-5 \pm 11$\,deg.

\textbf{WASP-44b} is a typical hot Jupiter orbiting a G8 host star on a 2.4-day orbit \citep{and12}. A uniform prior on $\nu\,\sin i_*$ resulted in $\lambda =-20^{+31}_{-34}$\,deg and $\nu\,\sin i_* = 0.8\pm0.5$ km/s; we found the amplitude of the R-M effect signal to be considerably weaker (by factor of 2) compared to the reported value $\nu\,\sin i_* = 3.2\pm0.9$ km/s. After stacking HARPS data we derive $\nu\,\sin i_*$= 1.5 $\pm$ 0.5 km/s with spectral synthesis using the iSpec \citep{cua14,blanco19} software. When applying a Gaussian prior around this value, we infer a prograde orbit of WASP-44b with $\lambda =-18^{+20}_{-26}$\,deg. We adopt the latter as our best estimate for $\lambda$.

\textbf{WASP-45 Ab} is a hot Jupiter orbiting a K2 host star on a short 3.1-day orbit. The host star has a seven magnitudes fainter companion, WASP-45 B, with a separation of \(4\arcsecond\).3 and a position angle of 317$^\circ$, corresponding to a M dwarf companion with a projected separation of 940 au. We infer an aligned orbit of WASP-45 Ab with $\lambda =13^{+15}_{-10}$\,deg.

\textbf{WASP-54 Ab} is an inflated Jupiter orbiting a F9 host star on a 3.7-day orbit \citep{fae13}. The host star has a less massive companion, WASP-54 B, with a separation of \(5\arcsecond\).7 and a position angle of 115.9$^\circ$, corresponding to a projected separation of 1440 au. We infer an aligned orbit of WASP-54 Ab with $\lambda =12 \pm 10$\,deg.

\textbf{WASP-91b} is a warm Jupiter orbiting a K3 host star on a 2.8-day orbit \citep{and17}. We infer an aligned orbit of WASP-91b with $\lambda =-11 \pm 20$\,deg.

\textbf{WASP-99 b} is a typical hot Jupiter. It orbits its F8 host star on a 5.8-day orbit \citep{hel14}. Due to the long transit duration, the egress was not observed. This incomplete transit coverage resulted in a broad posterior when using uniform priors on $\nu\,\sin i_*$
yielding $\lambda =2^{+27}_{-36}$\,deg. When applying Gaussian priors on the $\nu\,\sin i_*$, we infer an aligned orbit with $\lambda =-1^{+16}_{-18}$\,deg. We adopt this as our preferred solution.

\textbf{WASP-129 Ab} is a hot Jupiter orbiting its G1 host star on a 5.7-day orbit \citep{max16}.
\citet{mich24} reported the presence of a close binary companion with a separation of \(4\arcsecond\).34 and a position angle of 215.1$^\circ$, corresponding to a projected separation of 1370 au. Furthermore, based on the \textit{Gaia} mission photometry they identified the companion WASP-129B to be a white dwarf.  We derive a prograde orbit with $\lambda =10^{+6}_{-6}$\,deg.

\textbf{WASP-162b} is a massive warm Jupiter. It orbits its K0 host star on a 9.6-day, moderately eccentric ($e=0.434$), orbit \citep{hel19}. When applying uniform prior on $\nu\,\sin i_*$, we derive a prograde orbit with wide posterior $\lambda =7^{+48}_{-56}$\,deg as the $\nu\,\sin i_*$ wanders off to large values \citep[likely due to the degeneracy between $\nu\,\sin i_*$ and $\lambda$,][]{brow17} that are not compatible with the spectroscopic values. When we apply Gaussian priors on the $\nu\,\sin i_*$, we find that the orbit of WASP-162b is prograde, with $\lambda =1^{+32}_{-33}$\,deg. We adopt this as our preferred solution.

\textbf{Qatar-7b} is the only ultra-hot Jupiter in our sample. It orbits its F4 metal rich host star on a 2.0-day orbit \citep{als19}. Since the discovery paper provided wide priors from the photometry analysis and no follow-up of this system has been performed, we have analyzed unpublished TESS data to refine the parameters and present them in the Appendix. We find that the orbit of Qatar-7b is well aligned, with $\lambda =-9 \pm 11$\,deg.

\textbf{EPIC 219388192} is a system consisting of a transiting brown dwarf on 5.3-day orbit around its G1 star. Its mass puts it inside the brown dwarf desert \citep{gret06}. Furthermore, it is part of the Ruprecht 147 Open Cluster and has an astrometrically unresolved red dwarf binary companion on a 24 au orbit \citep{cur16,now17}. We find that the orbit of EPIC 219388192b is prograde, with $\lambda =16^{+16}_{-15}$\,deg.

The results of our fits are shown in Figs.\,\ref{f:35}--\ref{f:129}, while the MCMC results are shown in Figures placed on Zenodo\footnote{\url{https://zenodo.org/records/15259463}}. The derived values from the MCMC analysis are displayed in Table~\ref{tab:res3}.

\begin{deluxetable*}{lccccccc} 
\tablewidth{0pt} 
\tablecaption{MCMC analysis results for WASP-35b, WASP-44b, WASP-45 Ab, WASP-54 Ab, WASP-91b, WASP-99b, WASP-129 Ab, WASP-162b, Qatar-7b and EPIC 219388192b. $\mathcal{N}$ denotes priors with a normal distribution and $\mathcal{U}$ priors with a uniform distribution. \label{tab:res3}}
\tablehead{
\colhead{System} & \colhead{$T_c$ - 2450000} & \colhead{$\lambda$} & \colhead{$v\,\sin{i_*}$} & \colhead{$a/R_*$} & \colhead{Inc.} & \colhead{$\Gamma$} & \colhead{$K$} \\
\colhead{} & \colhead{[d]} & \colhead{[deg]} & \colhead{[km/s]} & \colhead{} & \colhead{[deg]} & \colhead{[km/s]} & \colhead{[km/s]}
}
\startdata
WASP-35b & $\mathcal{N}(T_0, 0.006)$ & $\mathcal{U}(-180, 180)$ & $\mathcal{U}(0, 50)$ & $\mathcal{N}(8.36, 0.13)$ & $\mathcal{N}(87.95, 0.33)$ & $\mathcal{N}(17.73, 0.20)$ & $\mathcal{N}(0.09, 0.05)$ \\
 & $8076.541 \pm 0.002$ & $-5 \pm 11$ & $2.1 \pm 0.2$ & $8.39 \pm 0.12$ & $87.93 \pm 0.32$ & $17.761 \pm 0.002$ & $0.08 \pm 0.01$ \\ \addlinespace 
\midrule 
WASP-44b & $\mathcal{N}(T_0, 0.006)$ & $\mathcal{U}(-180, 180)$ & $\mathcal{N}(1.5, 0.5)$ & $\mathcal{N}(8.05, 0.58)$ & $\mathcal{N}(86.0, 1.1)$ & $\mathcal{N}(-4.04, 0.20)$ & $\mathcal{N}(0.14, 0.10)$ \\
 & $8449.598 \pm 0.001$ & $-18^{+20}_{-26}$ & $1.2^{+0.4}_{-0.3}$ & $8.14 \pm 0.58$ & $85.7^{+1.1}_{-1.0}$ & $-4.004 \pm 0.004$ & $0.17 \pm 0.02$ \\ \addlinespace
\midrule
WASP-45 Ab & $\mathcal{N}(T_0, 0.006)$ & $\mathcal{U}(-180, 180)$ & $\mathcal{U}(0, 50)$ & $\mathcal{N}(8.58, 0.10)$ & $\mathcal{N}(84.5, 0.6)$ & $\mathcal{N}(4.55, 0.20)$ & $\mathcal{N}(0.15, 0.10)$ \\
 & $8426.673 \pm 0.001$ & $13^{+15}_{-10}$ & $1.1^{+0.5}_{-0.4}$ & $9.75^{+0.35}_{-0.36}$ & $84.98^{+0.29}_{-0.30}$ & $4.588 \pm 0.001$ & $0.130 \pm 0.007$ \\ \addlinespace
\midrule
WASP-54 Ab & $\mathcal{N}(T_0, 0.006)$ & $\mathcal{U}(-180, 180)$ & $\mathcal{U}(0, 50)$ & $\mathcal{N}(6.24, 0.17)$ & $\mathcal{N}(85.7, 0.5)$ & $\mathcal{N}(-3.13, 0.20)$ & $\mathcal{N}(0.07, 0.05)$ \\
 & $6057.622 \pm 0.001$ & $12 \pm 10$ & $2.0 \pm 0.2$ & $6.22 \pm 0.14$ & $85.95^{+0.50}_{-0.49}$ & $-3.124 \pm 0.002$ & $0.071 \pm 0.009$ \\ \addlinespace
\midrule
WASP-91b & $\mathcal{N}(T_0, 0.006)$ & $\mathcal{U}(-180, 180)$ & $\mathcal{U}(0, 50)$ & $\mathcal{N}(9.43, 0.20)$ & $\mathcal{N}(87.5, 0.3)$ & $\mathcal{N}(2.78, 0.20)$ & $\mathcal{N}(0.22, 0.20)$ \\
 & $8424.639 \pm 0.001$ & $-11.1 \pm 20$ & $0.5 \pm 0.2$ & $9.35^{+0.20}_{-0.19}$ & $87.64^{+0.28}_{-0.29}$ & $2.796 \pm 0.001$ & $0.207 \pm 0.007$ \\ \addlinespace
\midrule
WASP-99 b & $\mathcal{N}(T_0, 0.006)$ & $\mathcal{U}(-180, 180)$ & $\mathcal{N}(5.9, 0.4)$ & $\mathcal{N}(5.44, 0.10)$ & $\mathcal{N}(87.7, 1.0)$ & $\mathcal{N}(24.96, 0.20)$ & $\mathcal{N}(0.24, 0.20)$ \\
 & $6949.815 \pm 0.001$ & $-1^{+16}_{-18}$ & $5.6 \pm 0.3$ & $8.80^{+0.20}_{-0.22}$ & $89.25 \pm 0.83$ & $24.985 \pm 0.002$ & $0.24 \pm 0.02$ \\ \addlinespace
\midrule
WASP-129 Ab & $\mathcal{N}(T_0, 0.006)$ & $\mathcal{U}(-180, 180)$ & $\mathcal{U}(0, 50)$ & $\mathcal{N}(15.15, 0.39)$ & $\mathcal{N}(87.7, 0.2)$ & $\mathcal{N}(21.99, 0.20)$ & $\mathcal{N}(0.11, 0.10)$ \\
 & $7136.653 \pm 0.001$ & $9 \pm 6$ & $2.6 \pm 0.4$ & $15.09 \pm 0.34$ & $87.74 \pm 0.17$ & $21.988 \pm 0.002$ & $0.12 \pm 0.01$ \\ \addlinespace
\midrule
WASP-162b & $\mathcal{N}(T_0, 0.006)$ & $\mathcal{U}(-180, 180)$ & $\mathcal{N}(1.0, 0.5)$ & $\mathcal{N}(17.0, 0.5)$ & $\mathcal{N}(89.3, 0.5)$ & $\mathcal{N}(16.82, 0.20)$ & $\mathcal{N}(0.51, 0.20)$ \\
 & $8211.490 \pm 0.004$ & $1^{+31}_{-32}$ & $0.9 \pm 0.4$ & $17.04 \pm 0.47$ & $89.35 \pm 0.47$ & $17.079 \pm 0.003$ & $0.64 \pm 0.04$ \\ \addlinespace
\midrule
Qatar-7b & $\mathcal{N}(T_0, 0.006)$ & $\mathcal{U}(-180, 180)$ & $\mathcal{U}(0, 50)$ & $\mathcal{N}(4.89, 0.14)$ & $\mathcal{N}(85.65, 0.86)$ & $\mathcal{N}(-4.70, 0.20)$ & $\mathcal{N}(0.24, 0.20)$ \\
 & $8433.473 \pm 0.01$ & $-9 \pm 11$ & $10.7^{+1.6}_{-1.5}$ & $4.82 \pm 0.13$ & $85.31 \pm 0.83$ & $-4.73 \pm 0.01$ & $0.24 \pm 0.04$ \\ \addlinespace
\midrule
EPIC 219388192b & $\mathcal{N}(T_0, 0.006)$ & $\mathcal{U}(-180, 180)$ & $\mathcal{U}(0, 50)$ & $\mathcal{N}(12.96, 0.23)$ & $\mathcal{N}(88.53, 0.27)$ & $\mathcal{N}(47.0, 0.20)$ & $\mathcal{N}(4.20, 0.20)$ \\
 & $9431.625 \pm 0.001$ & $16^{+16}_{-15}$ & $4.9 \pm 0.8$ & $13.05 \pm 0.21$ & $88.47^{+0.26}_{-0.25}$ & $47.032 \pm 0.007$ & $4.05 \pm 0.04$ \\ \addlinespace
\enddata
\end{deluxetable*}

\section{Discussion}
\subsection{Stellar rotation and true spin-orbit angle $\psi$ }

By detecting variations in the light curve, we can determine the stellar rotation periods \citep[e.g.,][]{ska22}. This is crucial for estimating the stellar inclination $i_*$, which in turn is needed to determine the true spin-orbit angle $\psi$. The availability of this parameter in the literature is still limited (less than 50\% of planets with projected spin-orbit measurement have also their spin-orbit angle constrained) as it requires additional knowledge of the stellar inclination.  We attempted to derive the stellar rotation periods by analyzing ASAS-SN \textit{g}-band light curves \citep{ko17} using a Lomb-Scargle periodogram \citep{lomb76,sca82}.

We were able to detect a periodic modulation of EPIC 219388192 A with a period of $12.29 \pm 0.20$ days  (Figure on Zenodo), in  agreement with the rotational period of $12.6 \pm 2.1$ days reported by \citet{now17}. For the other targets, we were not able to detect any signal with high significance.

For EPIC 219388192 Ab, we are thus able to infer the true spin-orbit angle $\psi$ (i.e., the angle between the stellar spin-axis and the normal to the orbital plane) using the spherical law of cosines, $\cos \psi = \sin i_*\,\sin i\,\cos |\lambda|+\cos i_*\,\cos i$ with the approach suggested by \citet{mas20} accounting for the dependency between $v$ and $v\,\sin{i_*}$. We obtain $\psi = $25$^{+11}_{-14}$\,deg for EPIC 219388192 Ab, showing a prograde orbit with a slight but not significant misalignment.

\subsection{Dynamical timescales}

To better understand the evolution of the studied systems and the relevance of the derived parameters, we compare the estimated systems' ages with relevant dynamical timescales, starting with the circularization timescale, $\tau_{\text{cir}}$.

We use the following formula from \citet{adam06}:

\begin{equation}
 \tau_{\text{cir}} \approx \frac{4 Q'_P}{63} \left( \frac{a^3}{G M_*} \right)^{1/2} \frac{m_P}{M_*} \left( \frac{a}{R_P} \right)^5 (1 - e^2)^{13/2} [F(e^2)]^{-1} \label{eq:tau_cir}, 
\end{equation}
where $Q'_P$ is the tidal quality factor of the planet, which we assume to be $Q'_P = 10^6$, $a$ is the semi-major axis of the planet's orbit, $G$ is the universal gravitational constant, $M_*$ is the mass of the star, $m_P$ is the mass of the planet, $R_P$ is the radius of the planet, $e$ is the orbital eccentricity, and $F(e^2)$ is a correction factor accounting for non-zero eccentricity effects on the tidal interaction. 

We obtain $\tau_{\rm{cir}} \approx 4 \cdot 10^{10}$\,yr for WASP-162b, and $\tau_{\rm{cir}} \approx 2 \cdot 10^{11}$\,yr for EPIC 219388192b, the two planets showing eccentric orbits in our sample (see Table~\ref{tab:targets1}). As these timescales are significantly longer than the likely ages of the host stars (2-3 Gyrs), we do not expect these eccentric orbits to have circularized yet. The remaining planets in our sample have already circular orbits. We checked to see if these planets had acquired a small ($\approx 0.05$) eccentricity during their lifetimes, if the eccentric orbit would persist. We derive $\tau_{\rm{cir}} \approx 2 \cdot 10^{10}$\,yr and $\tau_{\rm{cir}} \approx 1 \cdot 10^{10}$\,yr for WASP-99 b and WASP-129\,Ab suggesting these orbits would not circularize. The remaining planets have circular orbits consistent with expected circularization timescales much shorter than their systems' ages ($\tau_{\rm{cir}} \lesssim 0.05$ Gyr); any initial eccentricity would likely have already been erased by tidal damping. Furthermore, the current circularity of these orbits also suggests that there is no significant ongoing excitation of eccentricity by an external companion, unlike the proposed scenario for WASP-140 system \citep{hel17}.

We also compare the system age to the tidal realignment timescale, which can reveal whether the current spin-orbit angle has been significantly altered by tides since formation. It provides only a limit on the possible realignment as the spin-orbit misalignment does not have to happen at the same time as the system's formation. For cool stars below the Kraft break \citep[][that is, with convective envelopes]{kraft67}, the tidal alignment timescale can be approximated \citep{alb12} as

\begin{equation}
\label{ce}
  \tau_{\rm{CE}} = \frac{10^{10}\,\rm{yr}}{\left( M_{\rm{p}}/M_{\rm{*}} \right)^2}  \left( \frac{a/R_{\rm{*}}}{40} \right)^6.
\end{equation}

\noindent
This allows us to compute the tidal alignment timescale for all the studied planets but Qatar-7, and derive the following timescales: $\tau_{\rm{Ce}} \approx 2 \cdot 10^{12}$\,yr for WASP-35b, $1 \cdot 10^{12}$\,yr for WASP-44b, $1 \cdot 10^{12}$\,yr for WASP-45 Ab, $2 \cdot 10^{12}$\,yr for WASP-54 Ab, $6 \cdot 10^{11}$\,yr for WASP-91b, $3 \cdot 10^{11}$\,yr for WASP-99b, $3 \cdot 10^{13}$\,yr for WASP-129b, $2 \cdot 10^{12}$\,yr for WASP-162b, and $1 \cdot 10^{10}$\,yr for EPIC 219388192b.

Qatar-7, with its spectral type F6, is classified as a hot star and thus lies above the Kraft break,  having a radiative envelope -- this structural difference has implications on the tidal forces and a different formula of the tidal alignment timescale \citep{alb12} should be used:

\begin{equation}
\label{ra}
\tau_{\rm{RA}} = \frac{5}{4} \cdot 10^9\, \text{yr}  \left( \frac{M_{\rm{p}}}{M_{\rm{*}}} \right)^{-2}  \left(1 + \frac{M_{\rm{p}}}{M_{\rm{*}}}\right)^{-5/6}  \left( \frac{a / R_{\star}}{6} \right)^{17/2},
\end{equation}

\noindent
leading to $\tau_{\rm{RA}} \approx 10^{14}$\,yr for Qatar-7b.

From the derived tidal alignment timescales of the targets, we can conclude that the current spin-orbit angles were likely not significantly changed from their initial values through tidal forces.
As we discuss in the following subsections, for WASP-162 and EPIC 219388192b due to moderate values of eccentricities, several migration mechanisms remain plausible. The low values of the spin-orbit angle of the remaining planets together with circular orbits are consistent with quiet disc migration \citep{bar14}.

\subsection{Spin-orbit angle of warm eccentric Jupiters}

Warm Jupiters show several differences compared to Hot Jupiters: \citet{hua16} noted that Warm Jupiters tend to display an overabundance of close companions and proposed there exist two populations with distinct formation mechanisms. \citet{wang24} suggested single-star Warm Jupiters tend to be more aligned compared to Hot Jupiters hinting at a more dynamical past of their hot counterparts. Warm Jupiters on eccentric orbits (dynamically hot systems) provide valuable insight into the various migration mechanisms as they are less affected by the tidal forces. Eccentric giant planets were often thought to be the result of high-eccentricity migration, potentially involving the Kozai-Lidov mechanism induced by a distant companion, which can simultaneously increase eccentricity and excite large spin-orbit angles \citep{fab07}. This mechanism was successfully confirmed, for example, in the highly eccentric (e=0.94) TIC 241249530 system with retrograde ($\lambda$=164 deg) planet and wide stellar companion ($\sim$ 1700 au) \citep{gup24}.  In addition, \citet{pet15} showed that secular interactions between multiple planets can excite high eccentricities without necessarily significantly affecting the spin-orbit angles (Coplanar high-eccentricity migration), unlike the ``traditional" Kozai-Lidov mechanism. This theoretical work has been lately supported by several cases of detection of eccentric warm Jupiters on aligned planets \citep{espi23,bie25}. 

Caution is needed before drawing firm conclusions from a handful of such systems, especially without complementary information like atmospheric composition. Alternative theories propose that eccentric planets on aligned orbits might also originate from disc migration itself, for example, through interactions with cavities or gaps within the protoplanetary disc that can excite planetary eccentricities \citep{deb21, li23}.

WASP-162b joins a handful of similar systems hosting a single transiting planet on a prograde and eccentric orbit with intriguing evolutionary past not well constrained by theoretical models. Future measurements on larger telescopes will be able to constrain the spin-orbit angle with better accuracy, allowing a more detailed description of the processes shaping the evolution of the system.

Long-term RV monitoring of WASP-162 and/or similar systems \citep[e.g., as shown for the TOI-677 system, which hosts a warm Jupiter and another object on a wider orbit,][]{seda23} might be an effective approach to break this degeneracy in evolutionary pathways.
The detection of another object in the WASP-162 system compatible with the coplanar high-eccentricity migration would provide evidence for this pathway \citep{zh24}. 

Another potential way to discriminate between disc migration and disc-free migration scenarios is by determining atmospheric elemental abundance ratios. Planets that underwent disc migration should have a different chemical composition than those migrating after the disc dispersal. However, due to the small atmospheric scale height of the WASP-162b, this planet is not particularly suited for atmospheric studies with the current generation of instruments.

\subsection{Spin-orbit angle of brown dwarfs}

The study of brown dwarfs, particularly those in transiting systems, offers crucial insights into the formation and evolution of objects bridging the gap between giant planets and stars \citep{hatz15}. The concept of the ``brown dwarf desert" \citep{gret06}—the observed scarcity of brown dwarf companions  within $\sim$ 3 au of main-sequence stars —remains a central theme, especially the particularly ``dry" region between 35–55 Jupiter masses for periods under 100 days. This desert has fueled discussions on formation mechanisms, with some proposing different pathways for brown dwarfs above and below this mass range \citep{ma14}. Several recent discoveries such as EPIC 212036875b, TOI-503, TOI-4776, TOI-5422 \citep{per19,sub20, zh25} reside within this debated desert region, showing that it might not be entirely empty. 

Measuring the spin-orbit alignment of brown dwarfs via the Rossiter-McLaughlin effect provides further dynamical clues about how this region is being populated, but such measurements have been relatively rare. Currently, only a handful of systems like CoRoT-3b \citep{tria09}, KELT-1b \citep{siv12}, WASP-30b \citep{triaud13}, and HATS-70b \citep{zhou19} have published spin-orbit measurements. The sample is slowly growing with recent additions like TOI-2533b \citep{fer24}, GPX-1b \citep{gia24}, LP 261-75C \citep{brad25}, and TOI-2119b \citep{doyle25} and all are consistent with an aligned orbit. 

Adding to this small but important sample, we infer the projected spin-orbit angle for EPIC 219388192b $\lambda =16^{+16}_{-15}$\,deg and $\psi = $25$^{+11}_{-14}$\,deg, consistent with a prograde orbit. This finding agrees with an emerging trend: brown dwarfs and other high mass-ratio companions $\frac{M_P}{M_*} > 2 \cdot10^{-3}$ tend to have low spin-orbit angles, even when orbiting hot stars where tidal realignment is expected to be inefficient \citep{rus24} and the population of Hot Jupiters shows a large diversity in the spin-orbit angle.

The specific configuration of the EPIC 219388192 system is also noteworthy. EPIC 219388192b's membership in the 2.5-3 Gyr old Ruprecht 147 cluster provides a crucial age benchmark, since deriving precise ages for isolated field stars older than 1 Gyr is rather difficult. It maintains a significant orbital eccentricity (e $\approx$ 0.19), consistent with its long tidal circularization timescale. Furthermore, it has a known M-dwarf outer companion on 24 au orbit, EPIC 219388192B, that could facilitate the high-eccentricity migration via the Kozai-Lidov mechanism raising the spin-orbit angle. For illustration, if we assume nearly circular orbit of the binary system, the Kozai-Lidov timescale \citep[using equation 27 from][]{naoz16} is around 300,000 years. This is much shorter than the system age, suggesting that the Kozai-Lidov mechanism could have had ample time to influence the system's architecture. 
The derived low spin-orbit angle for the brown dwarf, together with the tidal forces being inefficient, rather suggest disc migration for the system. However, because we had to make several assumptions (orbit of the binary, tidal quality factor) and the fact that the spin-orbit angle is not null, further research is needed to fully constrain the migration history.

\subsection{Spin-orbit angle in multi-star systems}

The spin-orbit alignment is being investigated in multi-star systems to understand how planets form and evolve in binary systems\footnote{In this work we talk only about s-type planets, that are
those that orbit around one star in a binary pair.} and how the presence of the binary companion affects the proto-planetary disc environment \citep{chris24, nea25}. This can be done by investigating the orbital orientation between exoplanets and wide-orbiting binary companions. \citet{chris24} reported the existence of an alignment between the orbits of small planets and binary systems with semimajor axes below 700 au. \citet{rice24} have studied the spin-orbit distribution of multi-star systems. They reported no clear correlation between spin-orbit misalignment and orbit-orbit misalignment. Furthermore, they reported the discovery of an overabundance of systems that are consistent with joint orbit-orbit and spin-orbit alignment.

In the sample presented here, three planets are in astrometrically resolved binary systems: WASP-45, WASP-54 and WASP-129. This allows us to compute the orbit-orbit angle $\gamma$ using \textit{Gaia} DR3 data:

\begin{equation}
 \cos \gamma = \frac{\vv{r} \cdot \vv{v}}{|\vv{r}| |\vv{v}|},
\end{equation}

where $\vv{r} \equiv [\Delta \alpha, \Delta \delta]$ and $\vv{v} \equiv [\Delta \mu^*_{\alpha}, \Delta \mu_{\delta}]$. Here, $\Delta \alpha$ and $\Delta \delta$ denote the positional differences between the primary and secondary in the right ascension (RA) and declination (Dec) directions, while $\Delta \mu^*_{\alpha}$ and $\Delta \mu_{\delta}$ are the proper motion differences in the RA and Dec directions. \textit{Gaia} DR3 reports a $\Delta \mu^*_{\alpha}$ value that has already the $ \cos \delta$ corrective factor in the RA direction incorporated\footnote{See the Proper motion in right ascension direction (PMRA) definition in the \textit{Gaia} DR3 source documentation at \url{https://gea.esac.esa.int/archive/documentation/GDR3/Gaia_archive/chap_datamodel/sec_dm_main_source_catalogue/ssec_dm_gaia_source.html}.} \citep{rice24}.

The angle \( \gamma \) quantifies the alignment between a binary system's relative position vector, $\vv{r}$, and relative velocity vector, $\vv{v}$. Values of \( \gamma \) near \( 0^\circ \) or \( 180^\circ \) suggest that the binary's motion is predominantly along our line of sight, characteristic of an edge-on orientation (orbit-orbit alignment). Conversely, \( \gamma \) values near \( 90^\circ \) indicate motion primarily perpendicular to our line of sight, aligning with a face-on orientation (orbit-orbit misalignment). However, interpreting \( \gamma \) for individual systems involves complexities: orbital eccentricity can cause variations in $\vv{r}$ and $\vv{v}$ that affect \( \gamma \). Furthermore, the planet-hosting star can be oriented within the sky-plane, hence accurately determining the alignment between the planetary orbit and the binary orbit is difficult. Therefore, while \( \gamma \) offers insights into the system's geometry, caution is necessary when interpreting its value due to these potential degeneracies. Significant improvements will be brought by the next \textit{Gaia} data releases as the current astrometry data are based solely on 34 months of observations \citep{gaia23}.

For WASP-45 Ab, we infer an aligned spin-orbit angle of $\lambda =13^{+15}_{-10}$\,deg. Furthermore, we derive edge-on orbit-orbit orientation with $\gamma =4 \pm 1$\,deg and a binary inclination of $168 \pm 14$\,deg for the WASP-45 system. For WASP-54 Ab we derive an aligned orbit with $\lambda =12 \pm 10$\,deg; additionally, we derive the orbit-orbit orientation of $\gamma =100 \pm 15$\,deg and a binary inclination of $40^{+22}_{-19}$\,deg for the WASP-54 system. For WASP-129 Ab, we infer a well-aligned spin-orbit angle of $\lambda =10^{+6}_{-6}$\,deg. Furthermore, we derive orbit-orbit orientation with $\gamma =112 \pm 1$\,deg and a binary inclination of $52^{+24}_{-23}$\,deg for the WASP-129 system.

WASP-129 system is only the second system after HAT-P-18 to have a planet with measured $\lambda$ and an astrometrically resolved white dwarf companion \citep{mug21an}. Both systems have their white dwarf companion at large separations (400-1500 au), yet their spin-orbit angle differs significantly \citep[aligned orbit for WASP-129\,Ab and retrograde orbit for HAT-P-18\,Ab][]{espo14} suggesting different evolutionary processes. Finally, the three systems in our sample do not fit into the sample identified by \citet{rice24} of fully aligned systems (showing spin-orbit, orbit-orbit alignment and edge-on binary configuration) and rather suggest more random orientations. Albeit, three new measurements do not affect the statistics in a significant way and measurements of additional planets in multiple stars systems are needed to understand their evolution and the interplay with the protoplanetary discs \citep{jo22,bar24}.

\section{Spin-orbit angles in the era of \textit{JWST} and \textit{Ariel}}

Several recent works have explored the distribution of measured exoplanet spin-orbit angles. Some studies proposed trends, such as a potential ``Preponderance of Perpendicular Planets," suggesting certain orbital configurations might be preferred \citep{alb21, att23}. In contrast, other analyses \citep[e.g.,][]{dong23, sie23} found the distribution to be largely consistent with isotropy, without strong evidence for preferred angles. Given the existing sample size of over 250 measurements, our contribution of 9 additional aligned systems is unlikely to significantly alter these population-level statistical results.

In this section, we explore the spin-orbit angle -- atmospheric connection of the already existing measurements. It has been recently stressed \citep{kirk24, pen24} that combining spin-orbit angle measurements with atmospheric characterization is crucial for a complete and robust understanding of an exoplanet's formation and evolution history. Spin-orbit measurements alone can be sometimes degenerate; for example, as discussed above, both coplanar high-eccentricity migration and certain disc-migration scenarios might produce planets on aligned, eccentric orbits. Atmospheric composition provides complementary information, potentially containing clues about formation location and accretion history. However, interpreting atmospheric elemental ratios (like C/O) is also complex, as processes like photochemistry, tidal heating, and vertical mixing can alter the atmospheric composition\footnote{Disequilibrium effects can alter the molecular abundances that are used to infer the elemental ratios, but these effects do not change the elemental ratios themselves.} over time \citep{kawa21}, partially obscuring the primordial signature. Therefore, combining both dynamical and compositional information is essential to break these degeneracies.

While the Hubble Space Telescope (HST) and James Webb Space Telescope (JWST) are powerful tools for atmospheric characterization, they operate largely via competed proposals and are not solely dedicated to exoplanet studies. In contrast, upcoming missions such as \textit{Ariel}, \textit{Twinkle}, \textit{Pandora} and others will be solely or heavily dedicated to the study of exoplanets and their atmospheres. The \textit{Ariel} mission is set to launch in late 2029. This mission aims to characterize the chemistry and thermodynamics of atmospheres in a homogeneous way for hundreds of exoplanets. \citet{ed22} have shown what a potential planet candidate list might look like, however, the final target list will be completed closer to the launch date. Our sample in this study consists of nine planets. All of them appear in the \textit{Ariel} Mission Candidate Sample (MCS). WASP-35b, and WASP-54 Ab are listed in Tier 3; WASP-44, WASP-45, WASP-91, WASP-99, Qatar-7 are listed in Tier 2 and WASP-129 and WASP-162 in Tier 1.

We examine the current status of spin-orbit measurements for planets already observed or approved for atmospheric characterization by HST and JWST. Our analysis includes approved targets up to HST cycle 32 and JWST targets from the first 4 cycles. We compiled a list of targets scheduled for transmission or emission spectroscopy using the TrExoLiSTS database \citep[][provides a compilation of transit, eclipse, and phase curve observations]{nik22}, supplemented with recent observing cycles. We then cross-matched this atmospheric target list with known projected spin-orbit angle measurements compiled from the TEPCat catalogue \citep{south11} and recent literature additions.

\begin{figure}[h]
\includegraphics[width=0.495\textwidth]{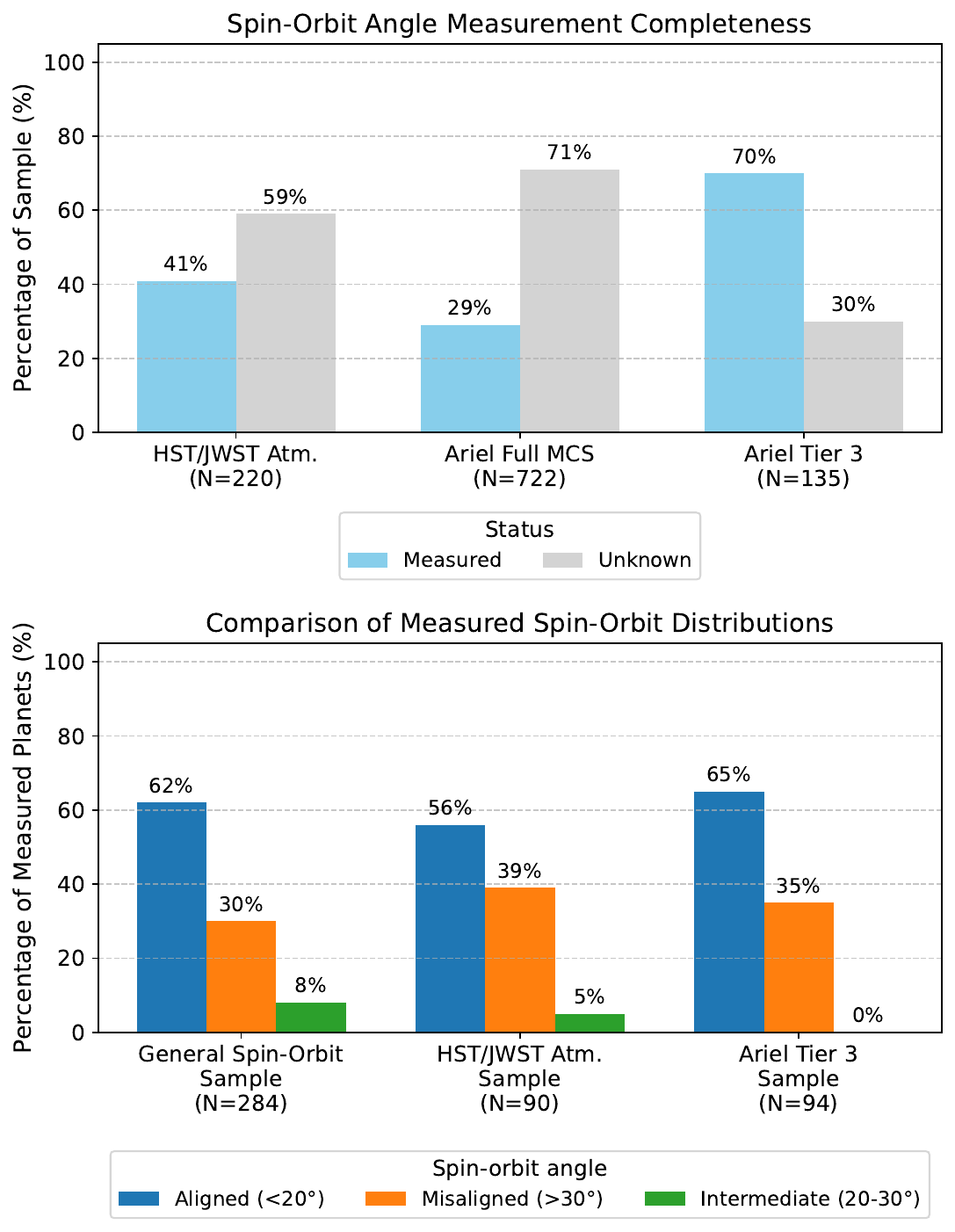}
\caption{Statistics of the planets with spin-orbit angle and atmospheric measurements; see the text for more details. (\textit{Top}) We show how many planets with \textit{JWST} and \textit{HST} measurements have their spin-orbit angle measured. We show the same for \textit{Ariel} Mission Candidate Sample (MCS) and \textit{Ariel} Tier 3 candidates. 
While only 41\% of \textit{HST}/\textit{JWST} atmospheric targets and 29\% of the full \textit{Ariel} sample have measured spin-orbit angles, the Ariel Tier 3 sample achieves 70\% completeness. (\textit{Bottom}) The total sample of planets with their spin-orbit angle measured shows 62\% aligned, 30\% misaligned, and 8\% intermediate cases. This is consistent with their distributions in the \textit{JWST}/\textit{HST} atmospheric sample and also \textit{Ariel} Tier 3 candidate list revealing no biases towards either population.}
\label{f:atmo}
\end{figure}

Our compilation resulted in 220 unique planets on the atmospheric target list and 284 planets with published projected\footnote{We used the projected spin-orbit angle rather than the true spin-orbit angle to increase the sample size.} spin-orbit angle measurements, with 90 planets common to both lists. 
Within the full list of 284 planets with spin-orbit measurements, the majority (62\%) have apparent aligned orbits (defined here as $|\lambda| < 20^\circ$), while 30\% have apparent misaligned orbits ($|\lambda| > 30^\circ$)\footnote{Planets with $20^\circ \le |\lambda| \le 30^\circ$ were excluded from these categories due to ambiguity.}. 
Turning to the 220 planets on the atmospheric target list, a significant fraction (59\%, or 130 planets) currently lack any published spin-orbit measurement. 
Of the 90 atmospheric targets with measured spin-orbit angle, 56\% (50 planets) are aligned ($|\lambda| < 20^\circ$) and 39\% (35 planets) are misaligned ($|\lambda| > 30^\circ$). 
While the proportion of misaligned planets among atmospheric targets with spin-orbit measurements (39\%) is slightly higher than in the overall spin-orbit sample (30\%), this difference is not statistically significant given the sample sizes. 
We cannot conclude that misaligned planets have been preferentially selected for atmospheric study thus far, e.g., due to tidal effects inflating the radius or the fact that observers prefer misaligned planets.
These results are illustrated in Figure~\ref{f:atmo}. 

We performed a similar analysis for potential \textit{Ariel} mission targets using the public MCS list\footnote{\url{https://github.com/arielmission-space/Mission_Candidate_Sample/tree/main/target_lists}}, specifically the ``$Ariel\_MCS\_Known\_2024-07-09.csv$" version containing 722 candidates. Overall, we found that only 29\% of these potential targets currently have their spin-orbit angle measurements published. As this MCS list is preliminary, we also focused on the 135 candidates currently designated as ``Tier 3". These are high-priority targets intended for detailed studies, potentially including atmospheric variability. The Tier sample will evolve as new planets are discovered (e.g., by \textit{TESS} and \textit{PLATO} missions).

For this Tier 3 subset, the completeness of spin-orbit measurements is much higher, at 70\%. 
Among these measured Tier 3 targets, 65\% are aligned ($|\lambda| < 20^\circ$) and 35\% are misaligned ($|\lambda| > 30^\circ$), proportions roughly compatible with the overall distribution in our compiled spin-orbit list. 
The results for the \textit{Ariel} sample analysis are also shown in Figure~\ref{f:atmo}.

In summary, the current sample of planets undergoing atmospheric characterization does not appear to be biased by their spin-orbit alignment. However, we highlight that a significant portion of these planets are missing measurements of their spin-orbit angles. The absence of this dynamical context limits our capacity to fully integrate these two approaches to understand the formation and evolution pathways, especially as we enter an era defined by high-precision data from \textit{JWST} and \textit{Ariel}.

\section{Summary}

The synergy between ground- and space-based observations is needed to study the evolution of planetary systems. Each approach is able to provide unique information and is also hindered by certain limitations that can be overcome by combining these approaches. In this work, we analyze transit spectroscopy data of ten systems on short orbits around FGK stars. Nine of these systems have a gas giant and one system consists of a transiting brown dwarf in a binary system. From the archival HARPS and HARPS-N data, we have analyzed the Rossiter-McLaughlin effect to infer the projected spin-orbit alignment of WASP-35b, WASP-44b, WASP-45 Ab, WASP-54 Ab, WASP-91b, WASP-99b, WASP-129 Ab, WASP-162 b, Qatar-7b and EPIC 219388192 b for the first time. We derive projected aligned orbits for all of the studied systems. Such measurements are consistent with disc migration. 
Furthermore, we have explored the completeness of the spin-orbit angle for planets with HST/JWST atmospheric measurements. We find no evidence of selection bias in existing atmospheric targets towards either aligned or misaligned population. We recommend future missions aiming for exoatmospheric measurements to take into account also the dynamical state of exoplanets to (a) avoid any biases in the sample selection, covering misaligned and aligned planets with the same detail and (b) to associate this key contextual information to composition results to more reliably identify any population-level trends.

\begin{acknowledgements} 
JZ and PK acknowledge support from GACR:22-30516K. JZ thanks the COST Action CA22133 PLANETS. AB is supported by the Italian Space Agency (ASI) with the Ariel grant n.~2021.5.HH.0.
\end{acknowledgements}

\software{ARoMEpy \citep{seda23}, matplotlib \citep{hun07}, exoplanet \citep{foreman21}, iSpec \citep{cua14}, NumPy \citep{harris2020array}, SciPy \citep{2020SciPy-NMeth}, astropy \citep{ast12,ast22}.}

\nocite{eno11,and12,fae13,and17,hel14,max16,hel19,als19,cur16,bal22,saha24,mac22,car19}
\bibliography{sample631}{}
\bibliographystyle{aasjournal}

\appendix

\section{\textit{TESS} photometry of Qatar-7} \label{apq7}

We analyzed photometric data of Qatar-7 (TIC 291685334) using observations from \textit{TESS} \citep{rick15}. The target was observed in two Sectors 57 and 84, during October 2022 and 2024 at a 120-second cadence. A total of 24 transits of Qatar-7b were captured, which were used to improve the planetary transit parameters.
We utilized the PDCSAP flux, which provides light curves corrected for instrumental systematics and stellar variability. To model the observed transits, we employed the \texttt{exoplanet} package \citep{foreman21}, a Bayesian framework for transit modeling.

\begin{figure}[h]
\includegraphics[width=0.5\textwidth]{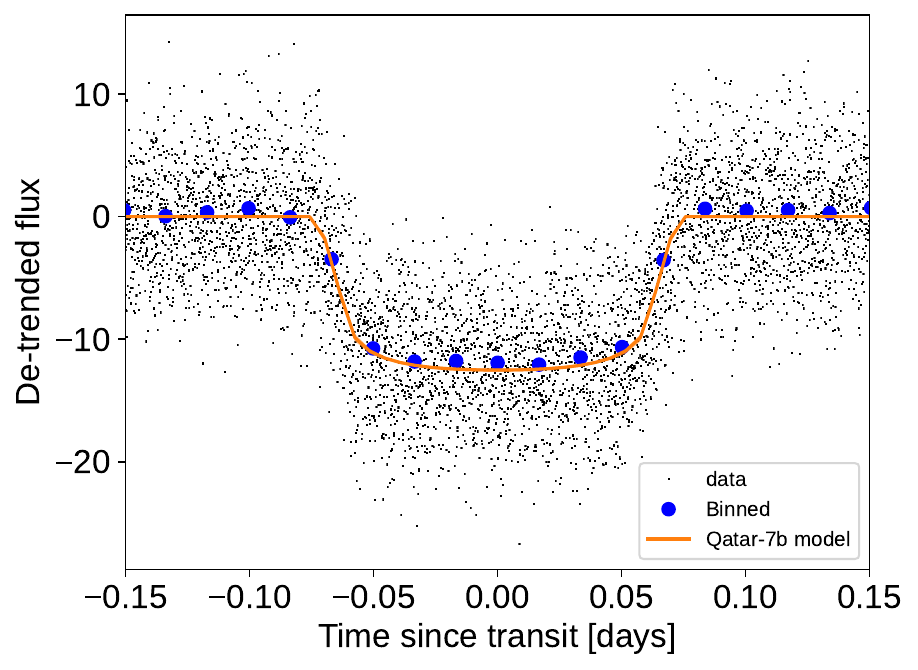}
\caption{Phase curve of Qatar-7b using \textit{TESS} data centered around the primary transit. Observed data are shown in black, binned data in blue and the obtained transit model in orange color.}
\label{fig:q7photo}
\end{figure}

The transit model assumed a Keplerian orbit and included a Gaussian Process (GP) to account for residual stellar variability. The stellar radius ($R_\star$) was constrained with a Gaussian prior centered on the value reported by \citet{als19}. Similarly, the orbital period ($P$) was modeled using a Gaussian prior based on previously determined ephemeris values. The transit depth ($\delta$) was also constrained with a Gaussian prior informed by earlier observations. A uniform prior between 0 and 1 was applied to the impact parameter ($b$). We adopted a quadratic limb-darkening law following \citet{kipping13}. The limb-darkening coefficients ($u_1, u_2$) were sampled with uniform priors constrained to the range between 0 and 1. The eccentricity $e$ and argument of periastron $\omega$ were fixed to values from \citet{als19}. To model residual stellar variability in the PDCSAP flux, we used a Matern-3/2 Gaussian Process kernel \citep{foreman17, foreman18}. We initialized the model using estimates for the orbital period, planetary radius, and reference transit time obtained from a Box Least Squares (BLS) periodogram analysis. The model parameters were first optimized using the maximum a posteriori (MAP) estimate. This was followed by Markov Chain Monte Carlo (MCMC) sampling using the No-U-Turn Sampler (NUTS). We performed 2000 tuning steps and sampled the posterior distributions with 7000 draws using three independent ensembles ensuring $\hat{R}$ value below 1.001 for all parameters. The obtained results are shown in Table \ref{tab:q7photo} and the phase curve shown in Fig.~\ref{fig:q7photo}.

\begin{table}[H]
\centering
\caption{Obtained parameters of Qatar-7b from fitting \textit{TESS} data.}
\label{tab:q7photo}
\begin{tabular}{l c}
\hline
\hline
\textbf{Parameter} & \textbf{Value} \\
\hline
\vspace{0.05cm}
$T_0$ (TJD) & $2853.85397^{+0.00051}_{-0.00051}$ \\
\vspace{0.05cm}
$P$ (days) & $2.0320211^{+0.0000016}_{-0.0000016}$ \\
\vspace{0.05cm}
$u_{\star, 0}$ & $0.14^{+0.11}_{-0.11}$ \\
\vspace{0.05cm}
$u_{\star, 1}$ & $0.44^{+0.24}_{-0.24}$ \\
\vspace{0.05cm}
$a / R_\star$ & $4.89^{+0.14}_{-0.14}$ \\
\vspace{0.05cm}
R$_{\rm{p}}$/R$_{\rm{s}}$ & $0.1048^{+0.0021}_{-0.0024}$ \\
\vspace{0.05cm}
$i$ (deg) & $85.65^{+0.84}_{-0.86}$ \\
\hline
\end{tabular}
\end{table}

\end{document}